\documentclass[%
reprint,
 amsmath,amssymb,
 aps,
 pra,
]{revtex4-1}

\usepackage{empheq}
\usepackage{graphicx}
\usepackage{epsfig}
\usepackage{epstopdf}
\usepackage{subfigure,color}
\usepackage{dcolumn}
\usepackage{bm}
\usepackage{amsmath,amssymb}
\usepackage{mathrsfs} \usepackage{comment}

\def\.{\cdot}

\def\##1{{\bf #1\mit}}
\def\_#1{{\bf #1\mit}}
\def\-#1{{\bf #1\mit}}
\def\=#1{\overline{\overline #1}}

\begin{document}

\preprint{APS/123-QED}

\title{Scattering from spheres made of time-varying and dispersive materials}
\author{G.~Ptitcyn$^{1,2}$}\email{These authors contributed equally.}
\author{A.~G.~Lamprianidis$^2$}\email{These authors contributed equally.}
\author{T.~Karamanos$^2$}\email{These authors contributed equally.}
\author{V.~S.~Asadchy$^3$}
\author{R.~Alaee$^2$}
\author{M.~M\"{u}ller$^2$}
\author{M.~Albooyeh$^4$}
\author{M.~S.~Mirmoosa$^1$}
\author{S.~Fan$^3$}
\author{S.~A.~Tretyakov$^1$}
\author{C.~Rockstuhl$^{2,5}$}
\affiliation{$^1$Department of Electronics and Nanoengineering, Aalto University, P.O.~Box 15500, FI-00076 Aalto, Finland\\$^2$ Institute of Theoretical Solid State Physics, Karlsruhe Institute of Technology, 76131 Karlsruhe, Germany\\$^3$Ginzton Laboratory and Department of Electrical Engineering, Stanford University, Stanford, California 94305, USA\\$^4$Mobix Labs Inc., 15420 Laguna Canyon, Irvine, California 92618, USA\\$^5$ Institute of Nanotechnology, Karlsruhe Institute of Technology, 76131 Karlsruhe, Germany}
\begin{abstract}
Exploring the interaction of light with time-varying media is an intellectual challenge that, in addition to fundamental aspects, provides a pathway to multiple promising  applications. Time modulation constitutes here a fundamental handle to control light on entirely different grounds. That holds particularly for complex systems simultaneously structured in space and time. However, a realistic description of time-varying materials requires considering their material dispersion. The combination thereof has barely been considered but is crucial since dispersion accompanies materials suitable for dynamic modulation. As a canonical scattering problem from which many general insights can be obtained, we develop and apply a self-consistent analytical theory of light scattering by a sphere made from a time-varying material exemplarily assumed to have a Lorentzian dispersion. We discuss the eigensolutions of Maxwell's equations in the bulk and present a dedicated Mie theory. The proposed theory is verified with full-wave simulations. We disclose effects such as energy transfer from the time-modulation subsystem to the electromagnetic field, amplifying carefully structured incident fields. Since many phenomena can be studied on analytical grounds with our formalism, it will be indispensable when exploring electromagnetic phenomena in time-varying and spatially structured finite objects of other geometries.  
\end{abstract}

\maketitle
\section{Introduction}
One of the most recent extensions of electromagnetics and optics is the concept of materials with time-varying properties. A time variation unlocks an additional degree of freedom in electromagnetic systems that tremendously increases the possibilities for controlling light-matter interactions~\cite{Engheta_4D,caloz2019spacetime,wang2020theory}. Temporal material modulations enable novel approaches to exceed conventional limitations~\cite{Our1,yu2009complete} and design efficient systems that realize unconventional functionalities~\cite{Our2,Halevi,pacheco2020temporal,pacheco2021temporal,lustig2018topological}. Historically, time-modulated structures were first studied when engineering radio-frequency antennas~\cite{jacob1954keying,wolff1957high,kummer1963ultra} to manipulate their bandwidth. In electronics, temporal modulations at twice the carrier frequency have been exploited in parametric amplifiers since the $19^\mathrm{th}$ century. Time-varying systems found applications from microwaves to optics. They led to discoveries of many intriguing phenomena such as magnetless nonreciprocity~\cite{yu2009complete,sounas2014angular,shi2017optical,dinc2017synchronized,fleury2018non,taravati2017nonreciprocal,Our2}, frequency conversion~\cite{cassedy1963dispersion,salary2018time}, amplification~\cite{holberg1966parametric,wang2018photonic,koutserimpas2018nonreciprocal}, Doppler shift~\cite{ramaccia2017doppler,ramaccia2019phase}, Fresnel drag~\cite{huidobro2019fresnel}, camouflage~\cite{liu2019time,wang2020spread}, breaking antenna performance  limits~\cite{hadad2016breaking}, temporal birefringence~\cite{akbarzadeh2018inverse,pacheco2021spatiotemporal}, temporal photonic crystals~\cite{lustig2018topological,sharabi2021disordered,shaltout2016photonic}, temporal discontinuities~\cite{zhou2020broadband}, power combiners~\cite{wang2021space}, light stopping and time reversal~\cite{yanik2004stopping,yanik2004time}, control of scattering and radiation~\cite{ptitcyn2019time,mirmoosa2020instantaneous}, enhanced wireless power transfer~\cite{jayathurathnage2021time}, control of absorption~\cite{mostafa2021coherent}, and more. However, the vast majority of prior contributions considered dispersionless materials. The absence of dispersion is synonymous with the assumption of an instantaneous (inertialess) response. This assumption is, generally speaking, non-physical, approximately holds  only for systems with very small variations over the time and/or negligible frequency dispersion. Examples of dispersionless materials whose properties can be modulated in time include lithium niobate (LiNBO$_3$)~\cite{barton2021wavefront} and  silicon~\cite{guo2019nonreciprocal} (in their transparency frequency regions). 
But the modulation depth of such dispersionless materials -- a parameter that determines the strength of the effects caused by temporal modulations -- is typically very low, being of the order of $10^{-4} - 10^{-3}$~\cite{williamson2020integrated}. In contrast, material candidates that allow large modulation depths of the permittivity, including electron plasmas~\cite{wang20193d} and aluminum-doped zinc~\cite{kinsey2015epsilon} and indium tin oxides~\cite{zhou2020broadband}, are usually strongly dispersive at the frequencies of interest (specifically, in  the epsilon-near-zero region, where the modulation depth is large). To our knowledge, only a few recent papers tackled this problem and studied the influence of frequency dispersion in  time-modulated materials~\cite{Our3,Soljacic,Engheta}. 

One of the development routes considers time modulation as an additional degree of freedom in spatially modulated structures such as metamaterials or metasurfaces. Constituents of such devices are meta-atoms with finite sizes in all three spatial dimensions. To the best of our knowledge, almost all previous contributions considered time-varying structures that are infinite in at least one spatial dimension, for instance, bulk media~\cite{solis2021time}, slabs~\cite{Soljacic}, and coatings of cylinders~\cite{salary2018time}. A couple of recent studies considered light scattering from finite-sized particles, a sphere and a conductive spherical shell, with time-varying properties~\cite{Stefanou:21,schab2021scattering}. However, accommodating dispersion in such models remains a challenge, and, therefore, it was set aside.

The problem of light scattering by a time-invariant and dispersive sphere was solved a century ago by Gustav Mie~\cite{mie1908beitrage}. An extension of this theory towards time-varying particles represents a solid initial step towards the design of time-varying metamaterials and metasurfaces. The study of light scattering from a sphere is also very instructive since many basic phenomena can be explored with analytical or  semi-analytical calculations. The gained insights can be applied to understand and explain the behavior of scatterers with a more complicated shape that require a full-wave numerical approach for their full exploration. Therefore, exploring the case of canonical objects, especially spheres, can be considered to be at the heart of scattering theory. An insight that one can obtain, for example, concerns the ability of spheres to support scattering resonances where either electric or magnetic multipole moments are driven into resonance.
During the last decade, passive metasurfaces made from scatterers supporting such Mie resonances have demonstrated a variety of novel optical phenomena~\cite{evlyukhin2010optical,garcia2011strong,geffrin2012magnetic}, and we envisage a substantial broadening of possible applications when time variations are considered as an additional degree of freedom in these systems. 

This paper extends the Mie theory to spheres made from a dispersive material with a periodically time-varying permittivity.
The findings of this paper are, nevertheless, applicable to an arbitrary aperiodic modulation in the limit of a very large period.
 First, we show how the dispersion relation of the eigenmodes of a homogeneous unbound dispersive medium transforms in the presence of temporal material modulations and analyze band structures of frequency-dispersive time crystals. Next, we introduce a field Ansatz in spherical coordinates for solving the scattering problem. The expressions for the T-matrix elements of  dynamic spheres are derived, and its power balance is analyzed. The analysis indicates a possibility to observe a negative absorption in the system, i.e., transfer of energy from time-varying matter to photons. This effect happens for an incident field carefully chosen in both its spatial and spectral distributions. Finally, we perform full-wave simulations of light scattering by a dispersive sphere via a finite-element time-domain method and find excellent agreement between the simulated and theoretical results. Based on the developed theory, one can further extend and generalize the analytical study of this paper towards structures with lower spatial dimensions such as infinite slabs and cylinders. Alternatively, the insights that are generated from these results can be useful to study arbitrarily shaped objects using solely full-wave simulations.
\section{Theoretical  analysis}
\begin{figure}
\centering
	\includegraphics[width=0.45 \textwidth]{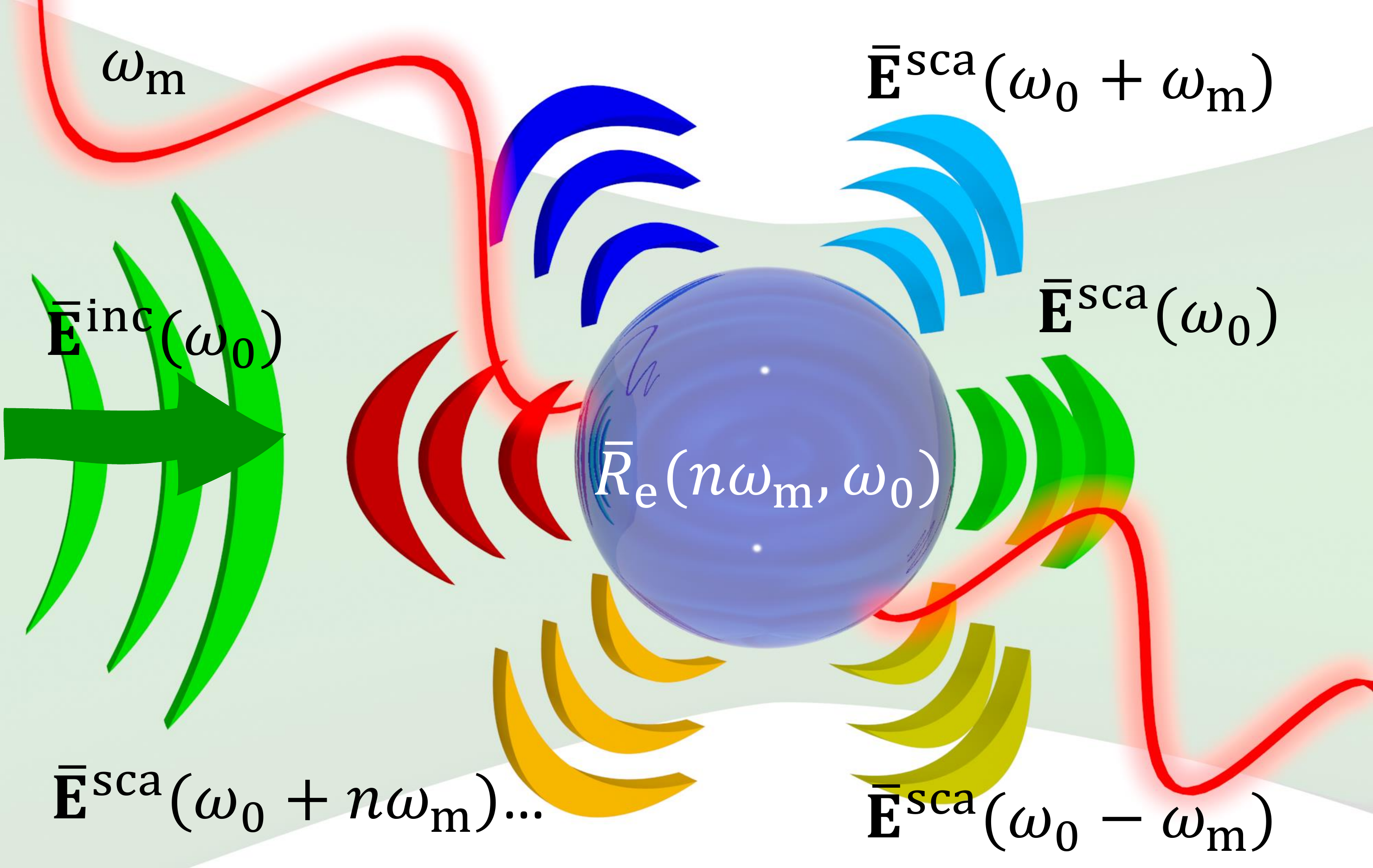}
	\caption{Illustration of the scattering of light by a sphere composed of a time-varying and dispersive medium.} 
	\label{fig_concept}
\end{figure}

In this section, we perform the electromagnetic analysis of the canonical problem of scattering by a sphere composed of a time-varying and dispersive medium, embedded in free space. The problem is illustrated in Fig.~\ref{fig_concept}. The section is organized into four subsections. First, we study the electromagnetic wave equation that governs the electromagnetic fields inside spatially homogeneous but time-varying and dispersive bulk media. Second, we discuss the response function of a medium modeled by a Lorentz-type oscillator equation with a time-varying bulk electron density. Then, we develop a  generalized Mie theory that treats the scattering problem of homogeneous spherical scatterers made of such time-varying and dispersive media. Finally, we present expressions for observable quantities such as the total scattered and absorbed power by such a scattering system.
\subsection{Unbounded time-varying media with frequency dispersion}
We begin by studying the equations that govern electromagnetic waves inside source-free homogeneous, isotropic, linear, non-magnetic, with only electric dipolar polarization response,  time-varying bulk media with temporal dispersion. In such media, we consider that Maxwell's equations are coupled with the following constitutive relations where the displacement vector $\mathbf{D}$ and the magnetic flux density $\textbf{B}$ are given by
\begin{subequations}\label{eq:cosntitutive2}
\begin{align}
\mathbf{D}(\mathbf{r},t)&=\hspace{3pt}\varepsilon_0\mathbf{E}(\mathbf{r},t)+\mathbf{P}(\mathbf{r},t),\\	\mathbf{B}(\mathbf{r},t)&=\mu_0\mathbf{H}(\mathbf{r},t),
	\end{align}
	\end{subequations}
	\noindent 
	where $\mathbf{E}$ and $\mathbf{H}$ are the electric and magnetic fields,  respectively, $\varepsilon_0$ is the electric permittivity of vacuum, and $\mu_0$ is the magnetic permeability of vacuum. The polarization vector  $\mathbf{P}(\mathbf{r},t)$ of the medium is given by~\cite{stepanov1976dielectric,Our3}
	\begin{eqnarray}
	\mathbf{P}(\mathbf{r},t)&=&\varepsilon_0\int\limits_{-\infty}^{+\infty}R_\mathrm{e}(t,t-\tau)\,\mathbf{E}(\mathbf{r},\tau)\,\mathrm{d}\tau\label{eq:displacement},
	\end{eqnarray}
where $R_\mathrm{e}(t,t-\tau)$ is the electric response function of the time-varying, dispersive medium. The response function $R_\mathrm{e}(t,t-\tau)$ expresses the polarization density $\mathbf{P}$ at time $t$ induced by an electric field impulse at time $\tau$. Equation~\eqref{eq:displacement} constitutes our Ansatz for the electric-field-driven polarization induced inside such a medium. It is important to note that this response function has the property $R_\mathrm{e}(t,t-\tau)=0$, for $t\leq\tau$, because of causality. Also, in the limiting case of non-time-varying media, the response function becomes invariant with respect to its first argument $t$. Moreover, in the limiting case of dispersionless media with an instantaneous response, the dependency of $R_\mathrm{e}(t,t-\tau)$ on its second argument $t-\tau$ is that of the Dirac delta distribution, $R_\mathrm{e}(t,t-\tau)=R'_\mathrm{e}(t)\delta(t-\tau)$. This last assumption of dispersionless media with an instantaneous response is found in several recent publications~\cite{Halevi,pacheco2020temporal,pacheco2021temporal,Stefanou:21}, but constitutes generally a physical assumption that is valid only in limited and approximate cases.
	
By making use of the Fourier transforms of the quantities involved, adopting the convention ${X}(t)=\frac{1}{\sqrt{2\pi}}\int_{-\infty}^{+\infty}\overline{X}(\omega)e^{-\mathrm{i}\omega t}\mathrm{d}\omega$, we will switch from the above time-space representation of the governing equations to the corresponding frequency-space representation. Note that bar sign represents frequency domain quantities in the Fourier transform. Therefore, Maxwell's equations read~\cite{aberg1995propagation,Our3,Soljacic}
\begin{subequations}
		\begin{flalign}
	\nabla\cdot\overline{\mathbf{E}}(\mathbf{r},\omega)&=-\hspace{-0.7mm}\int\limits_{-\infty}^{+\infty}{\overline{R}_\mathrm{e}}(\omega-\omega',\omega')\nabla\cdot\overline{\mathbf{E}}(\mathbf{r},\omega')\mathrm{d}\omega'\hspace{-0.3mm}, \label{eq:Max1}\\
    \hspace{0pt}\nabla\cdot\overline{\mathbf{H}}(\mathbf{r},\omega)&=0, \label{eq:Max2}\\
    \nabla\times\overline{\mathbf{E}}(\mathbf{r},\omega)&=\hspace{0pt}\mathrm{i}\omega\mu_0\overline{\mathbf{H}}(\mathbf{r},\omega), \label{eq:Max3}\\
    \nabla\times\overline{\mathbf{H}}(\mathbf{r},\omega)&=-\mathrm{i}\omega\varepsilon_0\overline{\mathbf{E}}(\mathbf{r},\omega)\nonumber\\
    &\hspace{3pt}-\mathrm{i}\omega\varepsilon_0\int\limits_{-\infty}^{+\infty}{\overline{R}_\mathrm{e}}(\omega-\omega',\omega')\overline{\mathbf{E}}(\mathbf{r},\omega')\mathrm{d}\omega'.\label{eq:Max4}
    \end{flalign}
\end{subequations}
\noindent The response function in frequency domain has been defined as the double Fourier transform ${\overline{R}_\mathrm{e}}(\omega-\omega',\omega')=\frac{1}{2\pi}\iint\limits_{-\infty}^{+\infty}R_\mathrm{e}(t,t-\tau)e^{\mathrm{i}(\omega t-\omega'\tau)}\mathrm{d}t\mathrm{d}\tau$. It gives the polarization denisty  $\overline{\mathbf{P}}(\omega)$ at frequency $\omega$ induced by an electric field impulse at frequency $\omega'$. Let us note that in the limiting case of non-time-varying media the response function in frequency domain ${\overline{R}_\mathrm{e}}(\omega-\omega',\omega')$ is a Dirac delta distribution with respect to its first argument and takes the following form: ${\overline{R}_\mathrm{e}}(\omega-\omega',\omega')=\delta(\omega-\omega')\chi(\omega)$, where it collapses into the usual electric susceptibility tensor. Moreover, in the limiting case of dispersionless media with instantaneous response, ${\overline{R}_\mathrm{e}}$ becomes invariant with respect to its second argument.
    
We move on by combining the last two equations to obtain the wave equation for the electric field:
        \begin{eqnarray}
    \nabla\times\nabla\times\overline{\mathbf{E}}(\mathbf{r},\omega)\hspace{165pt}&&\nonumber\\
    ={k}_0^2(\omega)\left[\overline{\mathbf{E}}(\mathbf{r},\omega)+\int\limits_{-\infty}^{+\infty}{\overline{R}_\mathrm{e}}(\omega-\omega',\omega')\overline{\mathbf{E}}(\mathbf{r},\omega')\mathrm{d}\omega'\right],\hspace{10pt}&& \label{eq:EwaveEq}
    \end{eqnarray}
\noindent where ${k}_0(\omega)=\omega\sqrt{\mu_0\varepsilon_0}=\omega/c_0$ is the wavenumber of free space. This wave equation can be simplified for dispersionless media as reported, e.g., in Ref.~\onlinecite{shi2016multi}. 

Next, we calculate the eigenfunctions that solve this homogeneous integro-differential equation, whose operator is non-diagonal in terms of the frequency $\omega$. These eigenfunctions are the fundamental solutions to these source-free Maxwell's equations. They are of paramount importance, as  general solutions induced by an arbitrary  source can always be written as a superposition of these fundamental solutions weighted with suitable amplitudes. 

To find these eigenfunctions, we use the method of separation of variables, and seek for solutions of the electric field $\overline{\mathbf{E}}(\mathbf{r},\omega)$ that have the following form: 
\begin{eqnarray}
\overline{\mathbf{E}}(\mathbf{r},\omega)=\int\mathcal{A}(\kappa)S_\kappa(\omega)\mathbf{F}_\kappa(\mathbf{r})\mathrm{d}\kappa,\label{eq:EigmodeExpans}
\end{eqnarray} 
\noindent where $\mathcal{A}(\kappa)$ is a complex amplitude. In this Ansatz, the dependency of the eigenfunctions on the spatial and frequency arguments is separated. By introducing the separation constant $\kappa^2$, we obtain the following set of coupled equations for the spatial part of the eigenfunctions, $\mathbf{F}_\kappa(\mathbf{r})$, and for the spectral part of the eigenfunctions, ${S}_\kappa(\omega)$:
\begin{subequations}
    \begin{eqnarray}
    \nabla\times\nabla\times\mathbf{F}_\kappa(\mathbf{r})=\kappa^2\mathbf{F}_\kappa(\mathbf{r})&&, \label{eq:EwaveEqSep1}\\
    {k}_0^2(\omega)\left[{S}_\kappa(\omega)+\int\limits_{-\infty}^{+\infty}{\overline{R}_\mathrm{e}}(\omega-\omega',\omega'){S}_\kappa(\omega')\mathrm{d}\omega'\right]\nonumber\\
    =\kappa^2{S}_\kappa(\omega). \label{eq:EwaveEqSep2}
    \end{eqnarray}
    \end{subequations}
Both of the above equations constitute themselves eigenvalue type of equations where $\kappa^2$ is the common eigenvalue. $\mathbf{F}_\kappa(\mathbf{r})$ is the corresponding eigenfunction of the differential operator of the first equation and ${S}_\kappa(\omega)$ is the corresponding eigenfunction of the integral operator of the second equation.  
The first equation~(\ref{eq:EwaveEqSep1}) for the spatial, vectorial profile of the eigenfunction $\mathbf{F}_\kappa(\mathbf{r})$, is an ordinary monochromatic electromagnetic wave equation with wavenumber $\kappa$. 
Depending on the coordinate system in which the solution is sought, the eigensolutions of the equation could be a set of plane waves, cylindrical waves, or spherical waves. As we wish to study electromagnetic scattering of light by a sphere, we choose spherical waves as eigensolutions, since they allow to apply the interface conditions needed.  At this point we could have picked plane waves or cylindrical waves, and proceed further in a similar way, if we were to study the interaction of light with a slab or an infinite cylinder, respectively. 

Such spherical-coordinate solutions to \eqref{eq:EwaveEqSep1} are known as vector spherical harmonics (VSHs). Details on their definition and properties can be found in Appendix~\ref{app:app1}. The VSHs are denoted here as $\mathbf{F}_{\alpha,\mu\nu\kappa}^{(\iota)}(\mathbf{r})$, where the index $``(\iota)"$ in the superscript takes the value ``(1)" to refer to regular VSHs finite at the origin $ \mathrm{r} =0$, or the value ``(3)" to refer to radiating VSHs that comply with the radiation condition at infinity. Hence, the former solutions describe standing waves, whereas the latter solutions describe scattered fields. Moreover, there is a set of four eigenvalues that characterize the VSHs. First, we have the index $\alpha$ that takes the values $\{\mathrm{M,N}\}$ to refer to transverse electric (TE) or transverse magnetic (TM) waves, i.e., multipoles of magnetic or electric type, respectively. Then, apart from the wavenumber $\kappa$, we also have the eigenvalues $\mu$ and $\nu$ with $\mu$ being the angular momentum along the $z$-axis and $\nu$ being the multipolar order of the VSH.

It is important to note that the spatial eigenfunctions $\mathbf{F}_\kappa(\mathbf{r})$ are solenoidal for $\kappa\neq0$, i.e., $\nabla\cdot\mathbf{F}_\kappa(\mathbf{r})=0$. For solutions with $\kappa=0$, it is straightforward to show that the magnetic field becomes irrotational, i.e., $\nabla\times\nabla\times\mathbf{F}_0(\mathbf{r})=0$. We then have a non-zero induced electric charge density distribution since the divergence of $\mathbf{F}_0(\mathbf{r})$ does not have to be zero. The above observation follows from (\ref{eq:Max1}), (\ref{eq:Max4}) and (\ref{eq:EigmodeExpans})--(\ref{eq:EwaveEqSep2}). Such implications that arise for $\kappa=0$ where the electric field ceases to be solenoidal are disregarded in the remainder of our analysis. This allows expanding the fields inside the time-varying scatterer only using the TE and TM spherical waves while avoiding the third multipolar family of longitudinal spherical waves~\cite{morse1953methods,fernandez2015exact,alaee2019exact}.

Let us now focus on the eigenvalue equation~(\ref{eq:EwaveEqSep2}). It is newly introduced by the time-variance of the medium and involves the spectral eigenfunction ${S}_\kappa(\omega)$. This eigenvalue equation plays the role of a dispersion relation of  time-varying systems. The important thing to notice here is that, due to the time-variance of the medium, the system is not translationally invariant in time, and we encounter a coupling among different frequency components. Equation~(\ref{eq:EwaveEqSep2}) governs this spectral coupling of electromagnetic field harmonics  inside the medium. In general, the equation has to be solved numerically by projecting it onto a Hilbert space $\mathscr{H}$ of finite dimensions. This leads to a finite linear system of equations whose eigenvalues $\kappa^2$ and corresponding eigenfunctions ${S}_\kappa(\omega)$ can be calculated numerically. Two assumptions have to be applied to make the system solvable.

First, we have to truncate the infinite spectrum into a finite spectral window. This will always numerically compromise the results. However, the truncation effect can be made arbitrarily small if the spectral window is sufficiently large compared to the spectral region of interest, since, usually, the spurious truncation effects will mainly affect the frequencies closer to the edges of the truncated spectral window. 

Second, we need to discretize the frequency $\omega$ in (\ref{eq:EwaveEqSep2}), which implies a time-periodic modulation of the medium. So, once we have that $R_\mathrm{e}(t,t-\tau)=R_\mathrm{e}(t+j T_\mathrm{m},t-\tau)$, with $T_\mathrm{m}$ being the modulation period and $j\in\mathbb{Z}$, the Fourier transform of the response function becomes discrete: ${\overline{R}_\mathrm{e}}(\omega-\omega',\omega')=\sum_j\delta(\omega-\omega'-j\omega_{\mathrm{m}}){\overline{R}'_\mathrm{e}}(\omega-\omega',\omega')$, with  $\omega_{\mathrm{m}}=2\pi/T_\mathrm{m}$ being the modulation frequency. Hence, for such a system with discrete translational symmetry in time, it is instructive to introduce a new eigenvalue, the Floquet frequency $\Omega$. This eigenvalue takes values within the frequency range $[0,\omega_{\mathrm{m}})$. 

Equation~(\ref{eq:EwaveEqSep2}) can now take the following discrete form:
\begin{subequations}
\begin{eqnarray}
    &&{k}_0^2(\Omega_j)\left[{S}_{\kappa}(\Omega_j)+\sum_{l=1}^{N_\Omega}{\overline{R}'_e}(\Omega_j-\Omega_l,\Omega_l){S}_{\kappa}(\Omega_l)\right]\hspace{40pt}\nonumber\\
    &&\hspace{150pt}=\kappa^2(\Omega){S}_{\kappa}(\Omega_j), \label{eq:EwaveEqSep2discrete}
    \end{eqnarray}
\text{where}
\begin{eqnarray}
\Omega_{j}=\Omega+(j+j_0)\,\omega_{\mathrm{m}}\label{eq:combfreq}
\end{eqnarray} 
\end{subequations}
with $j=1,2,\dots,N_\Omega$ and $j_0$ being an integer that we chose appropriately for the truncated spectral window of interest. $N_\Omega$ is the total number of frequencies of the discretized and truncated spectrum. 

We see that the Floquet frequency $\Omega$ constitutes an extra eigenvalue of our system. It characterizes an infinite periodic comb of frequencies passing through the frequency $\Omega$. Equation~(\ref{eq:combfreq}) gives the frequencies of such a spectral comb within an arbitrarily truncated spectral window. Due to the medium's periodic time modulation, only the frequencies contained in each such spectral comb are mutually coupled, constituting an independent system. Therefore, for each Floquet frequency $\Omega$, equation~(\ref{eq:EwaveEqSep2discrete}), repeated for all values of the index $j$, forms a linear system of equations that can be written in matrix form as 
\begin{eqnarray}
\hat{\mathbf{K}}(\Omega)\cdot\vec{{S}}_{\kappa}(\Omega)&=&\kappa^2(\Omega)\hspace{4pt}\vec{{S}}_{\kappa}(\Omega),\label{eq:EwaveEqSep2discreteMatrix}
\end{eqnarray}
where we have defined the vector:
\begin{eqnarray}
\vec{{S}}_{\kappa}(\Omega)&=&\left[{S}_{\kappa}(\Omega_1)\hspace{2pt}\cdots\hspace{2pt}{S}_{\kappa}(\Omega_{N_\Omega})\right]^\mathrm{T},\label{eq:spectralVec}
\end{eqnarray}
and the matrices:
\begin{eqnarray}
\hat{\mathbf{K}}(\Omega)&=&\hat{\mathbf{k}}_0^2(\Omega)\cdot\left[\hat{\mathbf{I}}+\hat{\mathbf{R}}_{\mathrm{e}}(\Omega)\right],\label{eq:spectralSysMatDef}\\
\hat{\mathbf{k}}_0(\Omega)&=&\mathrm{diag}\left[{k}_0(\Omega_1)\hspace{2pt}\cdots\hspace{2pt}{k}_0(\Omega_{N_\Omega})\right],\label{eq:k0Mat}
\end{eqnarray}
with $\hat{\mathbf{I}}$ being the identity matrix and with the $j$-th-row-, $l$-th-column-element of the matrix $\hat{\mathbf{R}}_{\mathrm{e}}(\Omega)$ being equal to $\overline{R}'_\mathrm{e}(\Omega_j-\Omega_l,\Omega_l)$.

Consequently, in the eigenvalue equation~(\ref{eq:EwaveEqSep2discreteMatrix}), we end up with a matrix $\hat{\mathbf{K}}(\Omega)$ of finite dimensions  $N_\Omega\times N_\Omega$, whose $N_\Omega$ eigenvalues $\kappa_{i}^2(\Omega)$ and corresponding eigenvectors $\vec{{S}}_{\kappa_i}(\Omega)$ can be calculated numerically for each Floquet frequency $\Omega$~\footnote{Let us note that at the limit of $T_{\rm m}\rightarrow\infty$, $j_0\rightarrow-\infty$, $N_\Omega\rightarrow\infty$, we get the general case of time-varying media that are not periodically modulated, where the discrete set of eigenvalues $\kappa_{i}^2(\Omega)$ ends up being a continuum of eigenvalues $\kappa$ in the complex plane, and the corresponding eigenvectors $\vec{{S}}_(\kappa_i)(\Omega)$ end up being the original spectral eigenfunctions ${S}_\kappa(\omega)$.}. The eigenvalues and the corresponding eigenvectors are enumerated by the index $i=1,\dots,N_\Omega$.

Finally, the expansion of the fields in (\ref{eq:EigmodeExpans}) can now take the following form within the Hilbert space $\mathscr{H}$ of finite dimensions that we constructed for the case of such a periodically modulated, time-varying medium:
\begin{eqnarray}
\boxed{
\overline{\mathbf{E}}(\mathbf{r},\omega)=\int_0^{\omega_{\mathrm{m}}}\sum_{i,j=1}^{N_\Omega}\mathcal{A}_i(\Omega)\delta(\omega-\Omega_j){S}_{\kappa_i}(\Omega_j)\mathbf{F}_{\kappa_{i}}(\mathbf{r})\mathrm{d}\Omega,}\nonumber\\\label{eq:EigmodeExpansdiscrete}
\end{eqnarray} 
\noindent with $\mathcal{A}_i(\Omega)$ being complex amplitudes. The above equation constitutes our general Ansatz for the expansion of fields inside periodically modulated, time-varying media~\footnote{Note that the integral here over the Floquet frequency $\Omega$ is defined as $\lim\limits_{\omega'_m\rightarrow\omega_{\mathrm{m}}}\int_0^{\omega'_m}\mathrm{d}\Omega$, but we keep it like this for brevity}. 
\subsection{The response function of time-varying media} \label{subsection:respfunctheory}
In this subsection, we will consider a simple case of a time-varying medium and derive its response function $\overline{R}_\mathrm{e}(\omega-\omega',\omega')$. We consider a polarizable medium in which there exist bound polarizable electrons of a single species that live inside the potential well of a Lorentzian harmonic oscillator. Let $\mathbf{p}(\mathbf{r},t)$ be the induced dipole moment of a single bound electron driven by the electric field. It shall obey the following differential equation of motion:
\begin{eqnarray}
\left[\frac{\partial^2}{\partial t^2}+\gamma_n\frac{\partial}{\partial t}+\omega_n^2\right]\mathbf{p}(\mathbf{r},t)&=&\frac{e^2}{m_e}\mathbf{E}(\mathbf{r},t),\label{eq:LorentzOscPDE}
\end{eqnarray}
where $\gamma_n$ is the damping factor of the oscillator, $\omega_n$ is its resonance frequency, and $e,m_e$ are the charge and the effective mass of an electron, respectively. 
The solution to the above differential equation is given by the following convolution integral:
\begin{subequations}
\begin{eqnarray}
\mathbf{p}(\mathbf{r},t)=\int_{-\infty}^{+\infty}\alpha_\mathrm{e}(t-\tau)\mathbf{E}(\mathbf{r},\tau)\,\mathrm{d}\tau,\label{eq:dipmomconv}
\end{eqnarray}
\text{where}
\begin{eqnarray}
\hspace{-9mm}\alpha_\mathrm{e}(t)\hspace{-0.7mm}&=&\hspace{-0.7mm}\frac{1}{\sqrt{\omega_n^2-\frac{\gamma_n^2}{4}}}\hspace{-0.1mm}\frac{e^2}{m_e}\hspace{-0.1mm}H(t)e^{-\frac{\gamma_n}{2}t}\hspace{-0.5mm}\sin\hspace{-0.5mm}\left(\hspace{-0.9mm}{t\hspace{-0.1mm}\sqrt{\omega_n^2-\frac{\gamma_n^2}{4}}}\hspace{0pt}\hspace{-0mm}\right),\label{eq:polarizability}
\end{eqnarray}
\end{subequations}
is the electric polarizability kernel of the Lorentzian oscillator and $H(t)$ is the Heaviside step function \cite{Our3}. 

Now, let us consider that the bulk electron density of these electron species, $N(t)$, gets modulated in time.  The polarization density of the medium shall then be given by the following equation:
\begin{eqnarray}
\mathbf{P}(\mathbf{r},t)=\int_{-\infty}^{+\infty}\alpha_\mathrm{e}(t-\tau)N(\tau)\mathbf{E}(\mathbf{r},\tau)\,\mathrm{d}\tau.\label{eq:poldensconv}
\end{eqnarray}
This equation implies that the electric field at each moment $\tau$ excites only the \textit{available} number of electrons in unit volume $N(\tau)$ \footnote{ Let us note that this model implies the assumption that the excited electrons decay as time tends to infinity according to the electric polarizability kernel. Ref.~\onlinecite{Our3} proposes another model for the response function of the medium that is given by the formula $R_\mathrm{e}(t,t-\tau)=\alpha_\mathrm{e}(t-\tau)N(t)/\epsilon_0$. We would like to highlight that, even though there may be various ways to model the response function of the time-varying medium depending on the particular physical considerations that one would need to adopt, any kind of such phenomenological model can be directly embedded in an identical way inside the rest of the theoretical framework that is developed in this section.}. Moreover, the model assumes that the electrons oscillate inside a Lorentzian potential well that remains invariant even if the bulk electron density varies in time. However, one would expect, for example, that the damping factor becomes larger with increasing bulk electron density, due to a higher rate of electron collisions. In this work we will avoid such considerations for simplicity.

It is straightforward to show that the polarization density obeys then the following differential equation:
\begin{eqnarray}
\left[\frac{\partial^2}{\partial t^2}+\gamma_n\frac{\partial}{\partial t}+\omega_n^2\right]\mathbf{P}(\mathbf{r},t)&=&\frac{e^2}{m_e}N(t)\mathbf{E}(\mathbf{r},t).\label{eq:difEqPolDens}
\end{eqnarray}
Such a model for the polarization density has already been reported in Refs.~\onlinecite{Engheta,PhysRevLett.125.127403}.

Furthermore, comparing the above equation with the Ansatz of Eq.~(\ref{eq:displacement}), we get that the response function of such a medium is equal to \begin{eqnarray}
R_\mathrm{e}(t,t-\tau)&=&\frac{\alpha_\mathrm{e}(t-\tau)N(\tau)}{\epsilon_0},\label{eq:respfunctime}
\end{eqnarray} 
which, in frequency domain, takes the following form:
\begin{eqnarray}
\overline{R}_\mathrm{e}(\omega-\omega',\omega')&=&\frac{1}{\sqrt{2\pi}}\frac{e^2}{m_e\varepsilon_0}\hspace{2pt}\frac{\overline{N}(\omega-\omega')}{\omega_n^2-\omega^2-\mathrm{i}\gamma_n\omega}.\label{eq:respfunc}
\end{eqnarray}
A more general type of response function could be given by a superposition of such Lorentz harmonic oscillators and an additional Drude term in order to account for different electron species~\cite{8261889}. 

In general, for a periodically modulated bulk electron density $N(t)$, with the modulation frequency $\omega_{\mathrm{m}}$, we have $\overline{N}(\omega-\omega')=\sum_j\mathcal{N}_j\delta(\omega-\omega'-j\omega_{\mathrm{m}})$, with $j\in\mathbb{Z}$ and $\mathcal{N}_j$ being complex coefficients. For example, for our numerical demonstration in the next section, we will consider the particular case where the bulk electron density is harmonically modulated in time according to $N(t)=N_0\left[1+M_\mathrm{s}\mathrm{cos}(\omega_{\mathrm{m}} t)\right]$, where $N_0$ is the bulk electron density of the unmodulated medium, $M_\mathrm{s}$ is the modulation strength, taking values from 0 to 1, and $\omega_{\mathrm{m}}$ is the modulation frequency. In such a case, we find that $\overline{N}(\omega-\omega')=\sqrt{2\pi}N_0\delta(\omega-\omega')+M_\mathrm{s}\sqrt{2\pi}N_0\left[\delta(\omega-\omega'+\omega_{\mathrm{m}})+\delta(\omega-\omega'-\omega_{\mathrm{m}})\right]/2$. We will use the above formulas in the next section to numerically solve the eigenvalue problem of Eq.~(\ref{eq:EwaveEqSep2discreteMatrix}).
\subsection{Scattering by time-varying spheres with frequency dispersion}
Let us move on to the particular problem of electromagnetic scattering by a sphere composed of a time-varying and dispersive material. The sphere has a radius ${R}$, is embedded in free space and centered at the origin of the coordinate system. We begin by expanding the incident field in the following series of regular VSHs~\cite{mishchenko2002scattering}:
\begin{eqnarray}
\overline{\mathbf{E}}^{\mathrm{inc}}(\mathbf{r},\omega)=\sum_{\nu\mu,\alpha}\mathcal{A}^{\mathrm{inc}}_{\alpha,\mu\nu}(\omega)\mathbf{F}^{(1)}_{\alpha,\mu\nu \mathrm{k}_0}(\mathbf{r}),\label{eq:EincExpans}
\end{eqnarray} 
where the free-space wavenumber ${k}_0(\omega)$ is a function of frequency. More details about the above expansion of the incident field can be found in Appendix~\ref{app:app2}. Accordingly, the scattered field can be expanded in the following series of radiating VSHs~\footnote{Note that we define the wavenumber of free space as ${k}_0(\omega)=\omega/c_0$ instead of ${k}_0(\omega)=|\omega|/c_0$. This ensures that we can use the VSHs that involve the spherical Hankel functions of the 1st kind in order to refer to outgoing spherical waves also for negative frequencies. In the other case, when $ {k}_0(\omega)=|\omega|/c_0$, we would need to switch to VSHs that involve the spherical Hankel functions of the 2nd kind in order to refer to outgoing spherical waves for negative frequencies. Such an alternative representation would be an equivalent one, since the spherical Hankel functions of the 1st and 2nd kind have the symmetry $\mathrm{h}^{(1)}_\nu(-x)=(-1)^\nu\mathrm{h}^{(2)}_\nu(x)$.}:
\begin{equation}
\overline{\mathbf{E}}^{\mathrm{sca}}(\mathbf{r},\omega)=\sum_{\nu\mu,\alpha}\mathcal{A}^{\mathrm{sca}}_{\alpha,\mu\nu}(\omega)\mathbf{F}^{(3)}_{\alpha,\mu\nu \mathrm{k}_0}(\mathbf{r}).\label{eq:EscaExpans}
\end{equation}
Finally, according to the Ansatz of (\ref{eq:EigmodeExpansdiscrete}), the field induced inside the sphere is expanded over the following series of regular VSHs:
\begin{eqnarray}
&&\overline{\mathbf{E}}^{\mathrm{ind}}(\mathbf{r},\omega)=\nonumber\\
&&\int_0^{\omega_{\mathrm{m}}}\sum_{i,j=1}^{N_\Omega}\sum_{\nu\mu,\alpha}\mathcal{A}^{\mathrm{ind}}_{\alpha,\mu\nu i}(\Omega)\delta(\omega-\Omega_j){S}_{\kappa_i}(\Omega_j)\mathbf{F}^{(1)}_{\alpha,\mu\nu \kappa_{i}}(\mathbf{r})\mathrm{d}\Omega.\nonumber\\\label{eq:EindExpans}
\end{eqnarray} 
Let us highlight here that, in the formula above, the eigenvalues $\kappa_i$~\footnote{It should be noted that the eigenvalues that we calculate numerically are $\kappa_i^2(\Omega)$. For this expansion of the fields, we select the principle branch of the square root $\kappa_i(\Omega)=+\sqrt{\kappa_i^2(\Omega)}$. This $\kappa_i(\Omega)$ is the wavenumber that we use in the regular VSHs of the expansion. The choice of the branch of the square root here does not play a role, since for regular VSHs we have the symmetry $\mathbf{F}^{(1)}_{\alpha,\mu\nu -\kappa_{i}}(\mathbf{r})=(-1)^\nu\mathbf{F}^{(1)}_{\alpha,\mu\nu \kappa_{i}}(\mathbf{r})$, which follows from the respective symmetry of  spherical Bessel functions. Therefore, picking the other branch of the square root simply leads to an equivalent representation.} and the frequencies $\Omega_j$, are functions of the Floquet frequency $\Omega$. This dependence is dropped in our notation here for simplicity but implicitly always assumed. In comparison to the two previous expansions of the fields in free space [Eqs.~(\ref{eq:EincExpans},\ref{eq:EscaExpans})], one can see here how our Ansatz for the fields inside the time-varying sphere [Eq.~(\ref{eq:EindExpans})] changes according to Eq.~(\ref{eq:EigmodeExpansdiscrete}). Due to the time variance, there is no unique wavenumber corresponding to each frequency anymore. Instead, we have a bunch of wavenumbers corresponding to each comb of frequencies characterized by the Floquet frequency $\Omega$. 

The series expansions of the respective magnetic fields can be taken by making use of the Maxwell-Faraday equation~(\ref{eq:Max3}), together with the following property of VSHs: $\nabla\times\mathbf{F}^{(\iota)}_{\alpha,\mu\nu \kappa}(\mathbf{r})=\kappa\mathbf{F}^{(\iota)}_{\beta,\mu\nu \kappa}(\mathbf{r})$, where $\beta\neq\alpha$.

Now, solving this electromagnetic scattering problem within the defined finite-dimensional Hilbert space $\mathscr{H}$ means calculating the unknown complex amplitudes $\mathcal{A}^{\mathrm{sca}}_{\alpha,\mu\nu}(\omega)$, $\mathcal{A}^{\mathrm{ind}}_{\alpha,\mu\nu i}(\Omega)$ given the amplitudes $\mathcal{A}^{\mathrm{inc}}_{\alpha,\mu\nu}(\omega)$. This can be done by imposing the following interface conditions on the surface of the sphere:
\begin{subequations}
\begin{eqnarray}
\hspace{-9mm}\left.\hat{\mathbf{r}}\hspace{-0.5mm}\times\hspace{-0.5mm}\left[\overline{\mathbf{E}}^{\mathrm{ind}}(\mathbf{r},\omega)\hspace{-0.2mm}-\hspace{-0.2mm}\overline{\mathbf{E}}^{\mathrm{sca}}(\mathbf{r},\omega)\hspace{-0.2mm}-\hspace{-0.2mm}\overline{\mathbf{E}}^{\mathrm{inc}}(\mathbf{r},\omega)\right]\right|_{ \mathrm{r} =R}\hspace{-0.2mm}&=&\hspace{-0.2mm}0,\label{eq:BC1}\\
\hspace{-9mm}\left.\hat{\mathbf{r}}\hspace{-0.5mm}\times\hspace{-0.5mm}\left[\overline{\mathbf{H}}^{\mathrm{ind}}(\mathbf{r},\omega)\hspace{-0.2mm}-\hspace{-0.2mm}\overline{\mathbf{H}}^{\mathrm{sca}}(\mathbf{r},\omega)\hspace{-0.2mm}-\hspace{-0.2mm}\overline{\mathbf{H}}^{\mathrm{inc}}(\mathbf{r},\omega)\right]\right|_{ \mathrm{r} =R}\hspace{-0.2mm}&=&\hspace{-0.2mm}0,\label{eq:BC2}
\end{eqnarray}
\end{subequations}
that enforce the continuity of the tangential fields according to Maxwell's equations. Here, we need to make use of the following orthogonality property of the VSHs~\cite{Moneda:07}:
\begin{eqnarray}
&&\frac{\oint_{S_{R}}\left[\hat{\mathbf{r}}\times\mathbf{F}_{\alpha,\mu\nu\kappa}^{(\iota)}(\mathbf{r})\right]\cdot\mathbf{F}_{\alpha',-\mu'\nu'\kappa'}^{(\iota')}(\mathbf{r})\hspace{1pt}\mathrm{d}s}{(-1)^{\mu+\delta_{\alpha N}}{R}^2\hspace{1pt}z^{(\iota')}_{\alpha',\nu}(\kappa'{R})}\hspace{80pt}\nonumber\\ &&\hspace{116pt}=\delta_{\alpha'\beta}\delta_{\mu'\mu}\delta_{\nu'\nu}\hspace{3pt}z^{(\iota)}_{\alpha,\nu}(\kappa R),\label{eq:orthogprop}
\end{eqnarray}
where integration is done over the spherical surface $S_{R}$ of radius $R$, upon which we need to enforce the above interface conditions. $\delta_{ij}$ is the Kronecker delta, $\beta\neq\alpha$, and $z^{(\iota)}_{\alpha,\nu}(x)$ is the generalized spherical Bessel function defined in Appendix~\ref{app:app1}. Finally, by substituting Eqs.~(\ref{eq:EincExpans}-\ref{eq:EindExpans}) into Eqs.~(\ref{eq:BC1},\ref{eq:BC2}) and by making use of Eq.~(\ref{eq:orthogprop}), we end up with the following inhomogeneous system of equations to be solved:
\begin{widetext}
\begin{subequations}
\begin{eqnarray}
\sum_i\mathcal{A}^{\mathrm{ind}}_{\alpha,\mu\nu i}(\Omega){S}_{\kappa_i}(\Omega_j)z^{(1)}_{\alpha,\nu}(\kappa_i R )&=&
\mathcal{A}^{\mathrm{sca}}_{\alpha,\mu\nu}(\Omega_j)z^{(3)}_{\alpha,\nu}(x_j)+\mathcal{A}^{\mathrm{inc}}_{\alpha,\mu\nu}(\Omega_j)z^{(1)}_{\alpha,\nu}(x_j),\label{eq:syseq1}\\
\sum_i\kappa_i\mathcal{A}^{\mathrm{ind}}_{\alpha,\mu\nu i}(\Omega){S}_{\kappa_i}(\Omega_j) z^{(1)}_{\beta,\nu}(\kappa_i R )&=&
{k_0}(\Omega_j)\left[\mathcal{A}^{\mathrm{sca}}_{\alpha,\mu\nu}(\Omega_j)z^{(3)}_{\beta,\nu}(x_j)+\mathcal{A}^{\mathrm{inc}}_{\alpha,\mu\nu}(\Omega_j)z^{(1)}_{\beta,\nu}(x_j)\right],\label{eq:syseq2}
\end{eqnarray}
\end{subequations}
\end{widetext}
where $x_j={k}_0(\Omega_j)R$. Let us note that the last two equations are equivalent to Eqs.~(22,23) of Ref.~\onlinecite{Stefanou:21}.
By introducing the following definitions of column vectors
\begin{subequations}
\begin{eqnarray}
\vec{\mathcal{A}}^{\mathrm{inc}}_{\alpha,\mu\nu}(\Omega)&=&\left[\mathcal{A}^{\mathrm{inc}}_{\alpha,\mu\nu}(\Omega_1)\hspace{2pt}\cdots\hspace{2pt}\mathcal{A}^{\mathrm{inc}}_{\alpha,\mu\nu}(\Omega_{N_\Omega})\right]^T,\label{eq:vecmatdef1}\\
\vec{\mathcal{A}}^{\mathrm{sca}}_{\alpha,\mu\nu}(\Omega)&=&\left[\mathcal{A}^{\mathrm{sca}}_{\alpha,\mu\nu}(\Omega_1)\hspace{2pt}\cdots\hspace{2pt}\mathcal{A}^{\mathrm{sca}}_{\alpha,\mu\nu}(\Omega_{N_\Omega})\right]^T,\label{eq:vecmatdef2}\\
\vec{\mathcal{A}}^{\mathrm{ind}}_{\alpha,\mu\nu}(\Omega)&=&\left[\mathcal{A}^{\mathrm{ind}}_{\alpha,\mu\nu 1}(\Omega)\hspace{2pt}\cdots\hspace{2pt}\mathcal{A}^{\mathrm{ind}}_{\alpha,\mu\nu N_\Omega}(\Omega)\right]^T,\label{eq:vecmatdef3}
\end{eqnarray}
\end{subequations}

and matrices

\begin{subequations}
\begin{flalign}
\hspace{-1.9mm}\hat{\mathbf{S}}(\Omega)\hspace{-0.7mm}&=\hspace{-0.7mm}\left[\vec{{S}}_{\kappa_1}(\Omega)\,\cdots\,\vec{{S}}_{\kappa_{N_\Omega}}(\Omega)\right],\label{eq:vecmatdef4}\\
\hspace{-1.9mm}\hat{\boldsymbol{\kappa}}(\Omega)\hspace{-0.7mm}&=\hspace{-0.7mm}\mathrm{diag}\hspace{-0.7mm}\left[\kappa_1(\Omega)\,\cdots\,\kappa_{N_\Omega}(\Omega)\right],\label{eq:vecmatdef5}\\
\hspace{-1.9mm}\hat{\mathbf{Z}}_{\alpha,\nu}^{(\iota)}\hspace{-0.1mm}(\Omega)\hspace{-0.7mm}&=\hspace{-0.7mm}\mathrm{diag}\hspace{-0.7mm}\left[z^{(\iota)}_{\alpha,\nu}(\kappa_1(\Omega) R )\,\cdots\,z^{(\iota)}_{\alpha,\nu}(\kappa_{N_\Omega}(\Omega) R )\right]\hspace{-0mm},\label{eq:vecmatdef6}\\
\hspace{-1.9mm}\hat{\mathring{\mathbf{Z}}}_{\alpha,\nu}^{(\iota)}\hspace{-0.1mm}(\hspace{-0.1mm}\Omega\hspace{-0.1mm})\hspace{-0.7mm}&=\hspace{-0.7mm}\mathrm{diag}\hspace{-0.8mm}\left[\hspace{-0.3mm}z^{(\iota)}_{\alpha,\nu}(\hspace{-0.2mm}{k}_0(\hspace{-0.1mm}\Omega_1\hspace{-0.1mm}) R \hspace{-0.1mm})\,\cdots\, z^{(\iota)}_{\alpha,\nu}(\hspace{-0.2mm}{k}_0(\hspace{-0.1mm}\Omega_{N_\Omega}\hspace{-0.1mm}) R \hspace{-0.1mm})\hspace{-0.3mm}\right]\hspace{-0.9mm},\hspace{-0pt}\label{eq:vecmatdef7}
\end{flalign}
\end{subequations}
together with the definitions in Eqs.~(\ref{eq:spectralVec},\ref{eq:k0Mat}), we can rewrite the above set of equations in the following matrix form:
\begin{widetext}
\begin{eqnarray}
&&\begin{bmatrix}
\hat{\mathbf{S}}\cdot\hat{\mathbf{Z}}_{\alpha,\nu}^{(1)} &-\hat{\mathring{\mathbf{Z}}}_{\alpha,\nu}^{(3)}\\
\hat{\mathbf{S}}\cdot\hat{\boldsymbol{\kappa}}\cdot\hat{\mathbf{Z}}_{\beta,\nu}^{(1)}&-\hat{\mathbf{k}}_0\cdot\hat{\mathring{\mathbf{Z}}}_{\beta,\nu}^{(3)}
\end{bmatrix}\cdot
\begin{bmatrix}
\vec{\mathcal{A}}^{\mathrm{ind}}_{\alpha,\mu\nu}\\
\vec{\mathcal{A}}^{\mathrm{sca}}_{\alpha,\mu\nu}
\end{bmatrix}=\begin{bmatrix}
\hat{\mathring{\mathbf{Z}}}_{\alpha,\nu}^{(1)}&\hat{\mathbf{0}}\\
\hat{\mathbf{0}}&\hat{\mathbf{k}}_0\cdot\hat{\mathring{\mathbf{Z}}}_{\beta,\nu}^{(1)}
\end{bmatrix}\cdot\begin{bmatrix}
\vec{\mathcal{A}}^{\mathrm{inc}}_{\alpha,\mu\nu}\\
\vec{\mathcal{A}}^{\mathrm{inc}}_{\alpha,\mu\nu}
\end{bmatrix},\label{eq:matsys}
\end{eqnarray}
\end{widetext}
where $\hat{\mathbf{0}}$ is a matrix with dimensions $N_\Omega\times N_\Omega$ filled with zeros. In the above equation, the dependencies on the Floquet frequency $\Omega$ were dropped for simplicity. Let us introduce now the following T-matrix
\begin{eqnarray}
&&\hat{\mathbf{T}}_{\alpha,\nu}(\Omega)=\begin{bmatrix}
\hat{\mathbf{T}}_{\alpha,\nu}^{11}&\hat{\mathbf{T}}_{\alpha,\nu}^{12}\\
\hat{\mathbf{T}}_{\alpha,\nu}^{21}&\hat{\mathbf{T}}_{\alpha,\nu}^{22}
\end{bmatrix}\hspace{130pt} \nonumber \\
&&\hspace{15pt}=\begin{bmatrix}
\hat{\mathbf{S}}\cdot\hat{\mathbf{Z}}_{\alpha,\nu}^{(1)}&-\hat{\mathring{\mathbf{Z}}}_{\alpha,\nu}^{(3)}\\
\hat{\mathbf{k}}_0^{-1}\cdot\hat{\mathbf{S}}\cdot\hat{\boldsymbol{\kappa}}\cdot\hat{\mathbf{Z}}_{\beta,\nu}^{(1)}&-\hat{\mathring{\mathbf{Z}}}_{\beta,\nu}^{(3)}
\end{bmatrix}^{-1}\cdot
\begin{bmatrix}
\hat{\mathring{\mathbf{Z}}}_{\alpha,\nu}^{(1)}&\hat{\mathbf{0}}\\
\hat{\mathbf{0}}&\hat{\mathring{\mathbf{Z}}}_{\beta,\nu}^{(1)}
\end{bmatrix}.\label{eq:Tmat}
\end{eqnarray}
By introducing also the following two T-matrices with dimensions $N_\Omega\times N_\Omega$:
\begin{subequations}
\begin{eqnarray}
\hat{\mathbf{T}}_{\alpha,\nu}^{\mathrm{ind}}(\Omega)&=&\hat{\mathbf{T}}_{\alpha,\nu}^{11}+\hat{\mathbf{T}}_{\alpha,\nu}^{12},\label{eq:Tmatind}\\
\hat{\mathbf{T}}_{\alpha,\nu}^{\mathrm{sca}}(\Omega)&=&\hat{\mathbf{T}}_{\alpha,\nu}^{21}+\hat{\mathbf{T}}_{\alpha,\nu}^{22},\label{eq:Tmatsca}
\end{eqnarray}
\end{subequations}
we finally end up with the following expressions for the complex amplitudes of the induced and scattered fields as functions of the incident amplitudes
\begin{subequations}
\begin{empheq}[box=\fbox]{align}
\vec{\mathcal{A}}^{\mathrm{ind}}_{\alpha,\mu\nu}(\Omega)&=\hat{\mathbf{T}}_{\alpha,\nu}^{\mathrm{ind}}(\Omega)\cdot\vec{\mathcal{A}}^{\mathrm{inc}}_{\alpha,\mu\nu}(\Omega),\label{eq:ETind}\\
\vec{\mathcal{A}}^{\mathrm{sca}}_{\alpha,\mu\nu}(\Omega)&=\hat{\mathbf{T}}_{\alpha,\nu}^{\mathrm{sca}}(\Omega)\cdot\vec{\mathcal{A}}^{\mathrm{inc}}_{\alpha,\mu\nu}(\Omega).\label{eq:ETsca}
\end{empheq}
\end{subequations}
The last two equations solve the scattering problem that we studied. 

Finally, let us discuss some important symmetry properties of the above T-matrices, defined by the spatio-temporal symmetries of the corresponding scattering system that they represent. First of all, due to the fact that our scattering system is time-varying, we end up having a T-matrix that is non-diagonal with respect to the frequency $\omega$. This property implies an inelastic scattering process. In fact, for the specific case of a time-modulated system with discrete translational symmetry over time, i.e., with a modulation period $T_\mathrm{m}$, according to the Floquet theorem, we get a T-matrix that is block diagonal over frequency $\omega$, with each block involving a comb of frequencies characterized by the Floquet frequency $\Omega$ and a period of $\omega_{\mathrm{m}}=2\pi/T_\mathrm{m}$. This is the sole change that the structure of the T-matrix undergoes due to the introduced time-variance of the scattering system. The spatial symmetries of the system of the spherical scatterer continue to be exactly the same as in the stationary case. Since such a scattering system is rotationally invariant with respect to the $z$-axis, we have a T-matrix that is diagonal with respect to the eigenvalue $\mu$, the angular momentum along the $z$-axis. Actually, rotational invariance of the system along an arbitrary axis, due to its spherical symmetry, implies also a T-matrix that is diagonal with respect to the multipolar order $\nu$. Moreover, due to the point inversion invariance of such a scattering system, we end up having a T-matrix that is diagonal with respect to the eigenvalue $\alpha$, since the TE and TM VSHs with a fixed multipolar order $\nu\pmod{2}$ have an opposite parity symmetry. Scatterers of different, non-spherical geometry, would generally break those spatial symmmetry properties of their T-matrices.
\subsection{Observable scattering quantities}
In this subsection we will provide expressions for the scattered and absorbed energy by spherical time-modulated scatterers. Following Eq.~(5.18) of Ref.~\onlinecite{mishchenko2002scattering}, as well as our conventions for the Fourier transforms of the fields and the above expressions for the incident and scattered fields in terms of series of VSHs, as presented in Eqs.~(\ref{eq:EincExpans},\ref{eq:EscaExpans}), we can get the following expressions for the total scattered energy $W^{\mathrm{sca}}$:
\begin{subequations}
\begin{eqnarray}
{W}^{\mathrm{sca}}&=&\int_0^{\omega_{\mathrm{m}}}\sum_{j=1}^{N_\Omega}{P}^{\mathrm{sca}}(\Omega_j)\,\mathrm{d}\Omega\nonumber \\
&=&\int_0^{\omega_{\mathrm{m}}}\sum_{j=1}^{N_\Omega}\sum_{\nu\mu,\alpha}\frac{\left|\mathcal{A}^{\mathrm{sca}}_{\alpha,\mu\nu}(\Omega_j)\right|^2}{{Z}_0{k}_0^2(\Omega_j)}\,\mathrm{d}\Omega,
\label{eq:Wsca1}
\end{eqnarray}
\text{and for the total absorbed energy ${W}^{\mathrm{abs}}$:}
\begin{eqnarray}
{W}^{\mathrm{abs}}&=&\int_0^{\omega_{\mathrm{m}}}\sum_{j=1}^{N_\Omega}{P}^{\mathrm{abs}}(\Omega_j)\,\mathrm{d}\Omega \nonumber\\
=&-&\int_0^{\omega_{\mathrm{m}}}\sum_{j=1}^{N_\Omega}\sum_{\nu\mu,\alpha}\frac{\Re\left\{\left[\mathcal{A}^{\mathrm{inc}}_{\alpha,\mu\nu}(\Omega_j)\right]^*\mathcal{A}^{\mathrm{sca}}_{\alpha,\mu\nu}(\Omega_j)\right\}}{{Z}_0{k}_0^2(\Omega_j)}\,\mathrm{d}\Omega,\nonumber\\\label{eq:Wabs1}
\end{eqnarray}
\end{subequations}
where ${P}^{\mathrm{sca}}(\omega)$, ${P}^{\mathrm{abs}}(\omega)$ are the total scattered and absorbed powers, respectively, and ${Z}_0$ is the wave impedance of free space.

We can also reach alternative expressions for the total scattered and absorbed energies by performing a singular value decomposition~\cite{PhysRevA.95.053834} of the following matrices:
\begin{eqnarray}
&&\hat{\mathbf{k}}_0^{-1}(\Omega)\cdot\hat{\mathbf{T}}_{\alpha,\nu}^{\mathrm{sca}}(\Omega)\cdot\hat{\mathbf{k}}_0(\Omega)\hspace{135pt}\nonumber\\
&&\hspace{85pt}=\hat{\mathbf{U}}_{\alpha,\nu}(\Omega)\cdot\hat{\mathbf{\Sigma}}_{\alpha,\nu}(\Omega)\cdot\hat{\mathbf{V}}_{\alpha,\nu}^{\dagger}(\Omega),
\label{eq:SVD}
\end{eqnarray}
where $\mathbf{\Sigma}_{\alpha,\nu}(\Omega)$ are diagonal matrices that contain the singular values $\sigma_{\alpha,\nu s}(\Omega)$ of the decomposition, and $\hat{\mathbf{U}}_{\alpha,\nu}(\Omega)$, $\hat{\mathbf{V}}_{\alpha,\nu}(\Omega)$ are matrices whose columns contain the correspondent left- and right-singular vectors, $\vec{{u}}_{\alpha,\nu s}(\Omega)$, $\vec{{v}}_{\alpha,\nu s}(\Omega)$, respectively.  The right- and left-singular vectors contain the incident and scattered multipolar spectra of the singular modes of the time-varying scattering system. By expanding the following vectors on the full basis of the right-singular vectors:
\begin{eqnarray}
\hat{\mathbf{k}}_0^{-1}(\Omega)\cdot\vec{\mathcal{A}}^{\mathrm{inc}}_{\alpha,\mu\nu}(\Omega)&=&\sum_{s=1}^{N_\Omega}\mathcal{S}^{\mathrm{inc}}_{\alpha,\mu\nu s}(\Omega)\vec{{v}}_{\alpha,\nu s}(\Omega),\label{eq:incsingexpans}
\end{eqnarray}
where $\mathcal{S}^{\mathrm{inc}}_{\alpha,\mu\nu s}(\Omega)=\vec{{v}}^\dagger_{\alpha,\nu s}(\Omega)\cdot\vec{\mathcal{A}}^{\mathrm{inc}}_{\alpha,\mu\nu}(\Omega)$ are complex coefficients, we can arrive to the following alternative expressions for the total scattered and absorbed energies:
\begin{subequations}
\begin{eqnarray}
{W}^{\mathrm{sca}}&=&\frac{1}{{Z}_0}\int_0^{\omega_{\mathrm{m}}}\sum_{\nu\mu,\alpha}\sum_{s=1}^{N_\Omega}\sigma^2_{\alpha,\nu s}\left|\mathcal{S}^{\mathrm{inc}}_{\alpha,\mu\nu s}\right|^2\mathrm{d}\Omega,\label{eq:Wsca2}\\
{W}^{\mathrm{abs}}
&=&-\frac{1}{{Z}_0}\int_0^{\omega_{\mathrm{m}}}\mathrm{d}\Omega\sum_{\nu\mu,\alpha}\nonumber\\
&&\sum_{s=1}^{N_\Omega}\sigma_{\alpha,\nu s}\left|\mathcal{S}^{\mathrm{inc}}_{\alpha,\mu\nu s}\right|^2\left[\sigma_{\alpha,\nu s}+\Re\left\{\vec{{v}}^\dagger_{\alpha,\nu s}\cdot\vec{{u}}_{\alpha,\nu s}\right\}\right]+\nonumber\\
&&\sum_{\substack{s,s'=1 \\ s'\neq s}}^{N_\Omega}\sigma_{\alpha,\nu s'}\Re\left\{\left[\mathcal{S}^{\mathrm{inc}}_{\alpha,\mu\nu s}\right]^*\mathcal{S}^{\mathrm{inc}}_{\alpha,\mu\nu s'}\vec{{v}}^\dagger_{\alpha,\nu s}\cdot\vec{{u}}_{\alpha,\nu s'}\right\},\nonumber\\\label{eq:Wabs2}
\end{eqnarray}
\end{subequations}
where we dropped the dependence of the quantities on the Floquet frequency $\Omega$ for simplicity. The sum in the last row of the last equation for the absorbed energy corresponds to couplings among the singular modes.
\section{Numerical study and discussion}
In this section, we will demonstrate and discuss numerical results based on the theoretical approach that we developed in the previous section. The section is divided into three subsections. In the first subsection, we present our results regarding the bulk media dynamics of time-varying and dispersive media and discuss their main electromagnetic properties. In the second subsection, we study the properties of the scattering system of a homogeneous spherical scatterer composed of a time-varying and dispersive medium. Properties of the T-matrix characterizing the scattering system are presented. We also highlight the prototypical ability of this system to act as an active element by transferring energy from the external modulation of the medium to the radiated electromagnetic field, resulting in negative electromagnetic absorption. In the last subsection, we numerically compare the developed semi-analytical approach to full-wave optical simulations, highlighting the accuracy and efficiency of our method.
\subsection{Bulk media dynamics}
\begin{figure*}
	\includegraphics[width=17cm]{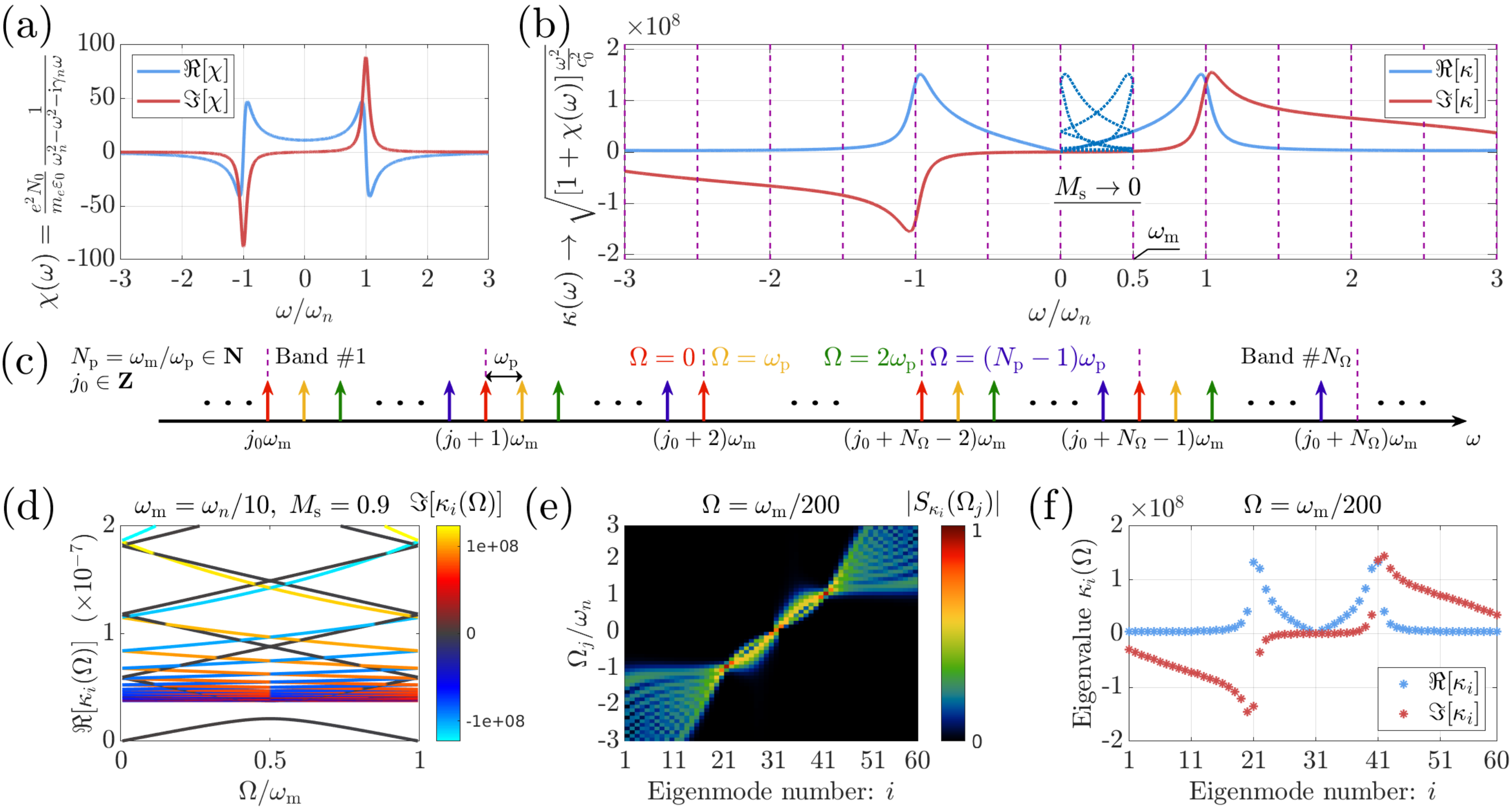}
	\caption{Bulk media dynamics: (a) Plot of the electric susceptibility of an unmodulated medium with dispersion corresponding to a single Lorentz oscillator. (b) Illustration of the folding of the band structure (dashed blue line) of the medium as we start modulating the bulk electron density of the Lorentz oscillator with frequency $\omega_\mathrm{m}$, in the limit of weak modulation strength $M_\mathrm{s}$. Solid lines represent the dispersion relation of the unmodulated medium. The dashed purple lines indicate the folding frequencies. (c) Illustration of the truncated, discrete spectrum of the response of a periodically modulated system excited by a periodic illumination. Spectral combs, characterized by the Floquet frequencies $\Omega$, constitute independent systems of coupled frequencies and are illustrated with different colors. Due to the periodic time-modulation of the system, there is coupling among the frequencies of each such spectral comb. (d) Plot of the band structure diagram of a time-modulated medium with strong modulation strength. The opening of a band gap can be observed. (e) Plot of the spectral content of the eigenmodes $\left|{S}_{\kappa_i}(\Omega_j)\right|$ that correspond to the system of a single comb of frequencies. We can observe the spectral coupling that is introduced by the strong time modulation of the medium. (f) Plot of the eigenvalues, i.e., the wavenumbers $\kappa_i$, that correspond to the eigenmodes presented in (e). } 
	\label{fig_2}
\end{figure*}
We begin by considering a dispersive medium described by a single Lorentz-type oscillator with natural resonance frequency $\omega_n$. The medium's electric susceptibility $\chi(\omega)$, in the non-time-varying case, is plotted in Fig.~\ref{fig_2}(a). The considered bulk electron density is $N_0=11\omega_n^2m_e\varepsilon_0/e^2$, and the damping factor of the oscillator is $\gamma_n=\omega_n/8$.

Then, we study the effect of the introduced time-variance on such dispersive medium. To do that, we assume that the bulk electron density of the Lorentz oscillator varies harmonically in time according to $N(t)=N_0\left[1+M_\mathrm{s}\cos(\omega_{\mathrm{m}} t)\right]$ and the corresponding response function of the medium is given by Eq.~(\ref{eq:respfunc}) according to the analysis of Sec.~\ref{subsection:respfunctheory}. We use this time modulation to study the eigenvalues, i.e., the supported wavenumbers $\kappa(\Omega)$, and the corresponding eigenvectors $\vec{S}_\kappa(\Omega)$, i.e., the corresponding spectral eigenfunctions, of such a system as described by Eq.~(\ref{eq:EwaveEqSep2discreteMatrix}). 

With the purpose to ease the understanding, let us discuss initially what happens in the limiting case of very small modulation strengths $M_\mathrm{s}\rightarrow0$. In analogy to periodically spatially modulated materials, this case would correspond to an empty-lattice approximation, i.e., practically there is no modulation but the periodicity is still introduced. This assumption allows us to work with the analytical dispersion relation while observing the onset of a band structure. This approach facilitates the understanding of the further results. 

In this case, the integral operator in Eq.~(\ref{eq:EwaveEqSep2}) remains predominantly diagonal, with very small off-diagonal terms. This property  indicates a weak spectral coupling among frequencies. This means that the spectral eigenfunctions ${S}_\kappa(\omega)$ tend to delta distributions. They tend to associate a unique wavenumber $\kappa$ to each frequency $\omega$, as is the case for the usual dispersion relation of non-time-varying media. Thus, we see that $\kappa(\omega)\rightarrow\sqrt{\left[1+\chi(\omega)\right]\omega^2/c_0^2}$, as $M_\mathrm{s}\rightarrow0$. We show in Fig.~\ref{fig_2}(b) how the band structure of such a system is formed by folding the wavenumbers $\kappa(\omega)$ into the fundamental spectral band. This spectral band ranges from the zero frequency to the modulation frequency of the medium $\omega_{\mathrm{m}}$, encompassing in this manner all the Floquet frequencies $\Omega$. The blue dashed lines show the band structure formed by folding the solid blue line within the fundamental spectral band that corresponds to a modulation frequency $\omega_{\mathrm{m}}=\omega_n/2$. The folding takes place periodically in frequencies denoted by the dashed purple lines.

We illustrate in Fig.~\ref{fig_2}(c) the truncated discrete spectrum of the response of a system periodically modulated with a frequency $\omega_{\mathrm{m}}$ and excited, also, by a periodic excitation. $\omega_{\mathrm{p}}=\omega_{\mathrm{m}}/N_\mathrm{p}$, with $N_\mathrm{p}\in\mathbb{N}$, is the frequency that corresponds to the superperiod of the combined periodicities of the modulation and excitation. We can see that the spectrum can be separated into a set of $N_\mathrm{p}$ finite combs of frequencies with a periodicity of $\omega_{\mathrm{m}}$. Each such comb of frequencies, indicated by a different color, corresponds to a different Floquet frequency $\Omega$, which varies between zero and $\omega_{\mathrm{m}}$. Due to the periodic time modulation of the medium, there is only coupling among the frequencies of each such spectral comb. Spectral combs of different Floquet frequencies $\Omega$ constitute independent systems and do not couple to each other. For numerical reasons, we have to truncate the spectrum. Here, a truncated spectral window of $N_\Omega$ bands is illustrated, which corresponds, also, to the number of frequencies of each such truncated spectral comb. The actual distribution of power among the frequencies of each comb can be decomposed over the full basis of eigenvectors $\vec{{S}}_{\kappa_i}(\Omega)$.

In general, what changes with the introduced periodic time modulation of the medium, is that now we have a set of wavenumbers $\kappa_i(\Omega)$ associated with each Floquet frequency $\Omega$, and with each wavenumber corresponding to a different, generally broadband, spectrum, given by the corresponding eigenvectors $\vec{{S}}_{\kappa_i}(\Omega)$. We solve the eigenvalue equation~(\ref{eq:EwaveEqSep2discreteMatrix}) and plot in Fig.~\ref{fig_2}(d)  the band structure, i.e., the set of eigenvalues/wavenumbers that correspond to each Floquet frequency $\Omega$, i.e., to each spectral comb. For the pertinent case we consider a modulation strength $M_{S}=0.9$ and a modulation frequency $\omega_{\mathrm{m}}=\omega_n/10$. The color of the line in this figure encodes the imaginary part of the eigenvalues. Generally, positive and negative imaginary values correspond to spectral eigenmodes with predominant spectral content over positive and negative frequencies, respectively. Let us note that the graph focuses on the region of eigenvalues with a small real part. We can see that, due to strong enough time modulation of the medium, we introduce a band gap in the lower band of the band structure~\cite{Halevi,wang2018photonic,lustig2018topological}. This is not possible for low modulation strengths as the effect is indicative of strong spectral coupling.

In Figs.~\ref{fig_2}(e,f) we plot, respectively, the eigenvectors and the corresponding eigenvalues of the subsystem with Floquet frequency $\Omega=\omega_{\mathrm{m}}/200$. The eigenmodes are ordered with respect to ascending eigenmode central frequency, which is defined as the following sum:  $\sum_j\Omega_j\left|{S}_{\kappa_i}(\Omega_j)\right|^2$. We can see in Fig.~\ref{fig_2}(e) that each eigenmode has a different spectral content distributed over the frequencies of the spectral comb characterized by the particular Floquet frequency $\Omega$. The spectral support becomes wider for eigenmodes that support high frequencies, whereas for the eigenmodes that predominantly support the frequencies $\Omega_j/\omega_n\approx0,\pm1$ it becomes minimally narrow. As discussed previously, the matrix with the eigenvectors plotted in Fig.~\ref{fig_2}(e) shall approach the identity matrix in the limit of $M_{S}\rightarrow0$. The degree of non-diagonality of the matrix $\hat{\mathbf{S}}(\Omega)$ is indicative of the strength of the spectral coupling within the time-varying system. In Fig.~\ref{fig_2}(f), we plot the sorted eigenvalues, i.e., the wavenumbers  associated with the corresponding eigenvectors as sorted in Fig.~\ref{fig_2}(e). Let us note the resemblance of Fig.~\ref{fig_2}(f) with the unmodulated case illustrated in  Fig.~\ref{fig_2}(b). The sorted wavenumbers of the strongly-modulated case are quite similar to those of the unmodulated case. However, we still have quite significant deviations, as it is indicated by the presence of the open band gap in Fig.~\ref{fig_2}(d).
\subsection{The scattering system of a time-varying and dispersive sphere}
\begin{figure}
	\includegraphics[width=8.6cm]{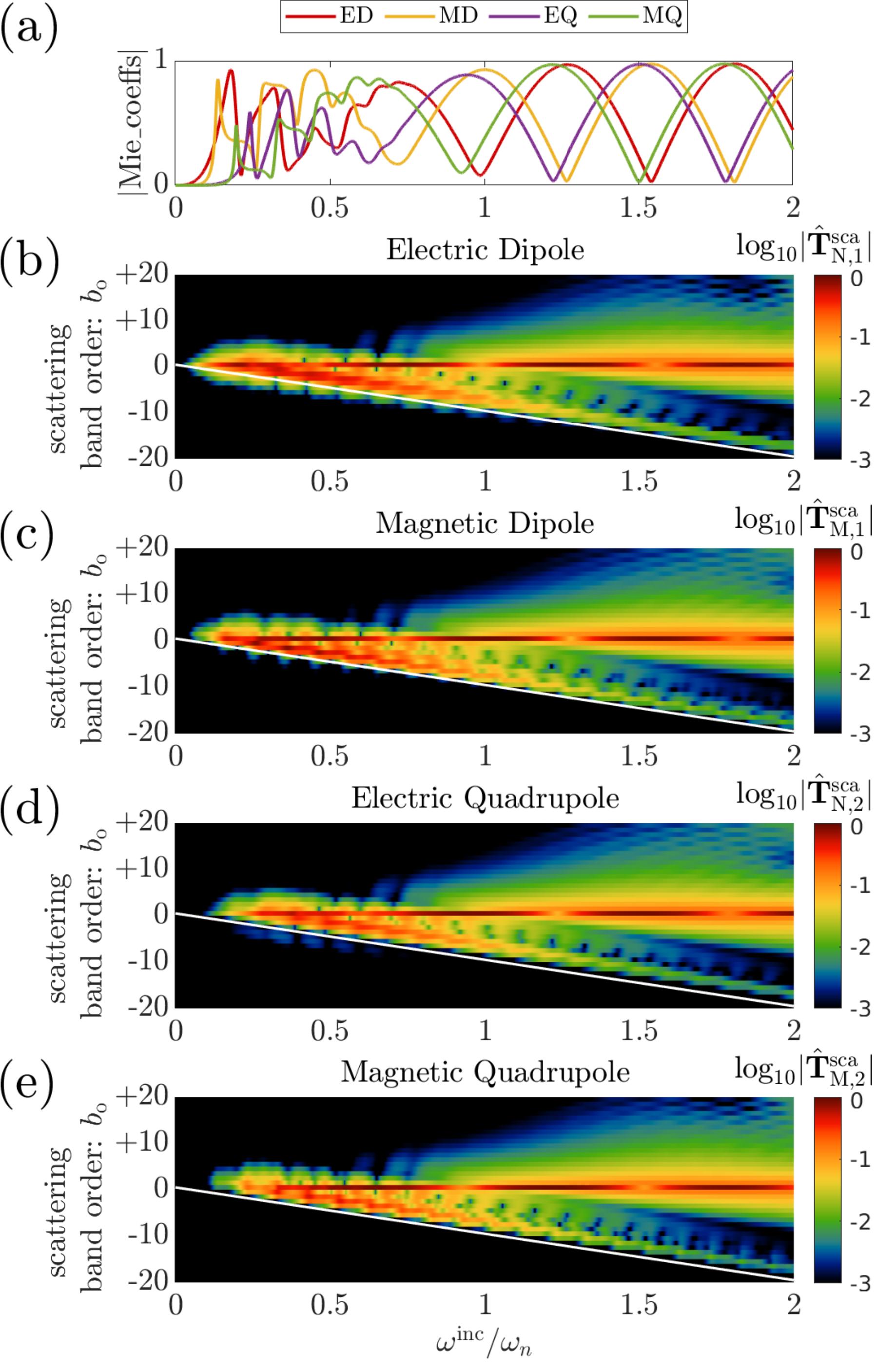}
	\caption{The scattering system of a time-varying and dispersive sphere: (a) Plot of the absolute value of the Mie coefficients that correspond to a sphere made of a dispersive medium without time modulation. (b-e) Plots of the T-matrix elements of a sphere with introduced strong time modulation, for different multipolar orders. Time modulation leads to an inelastic scattering process where there is spectral coupling among different input and output frequencies given by $\omega^{\mathrm{sca}}=\omega^{\mathrm{inc}}+b_\mathrm{o}\omega_{\mathrm{m}}$.} 
	\label{fig_3}
\end{figure}
\begin{figure*}
	\includegraphics[width=17cm]{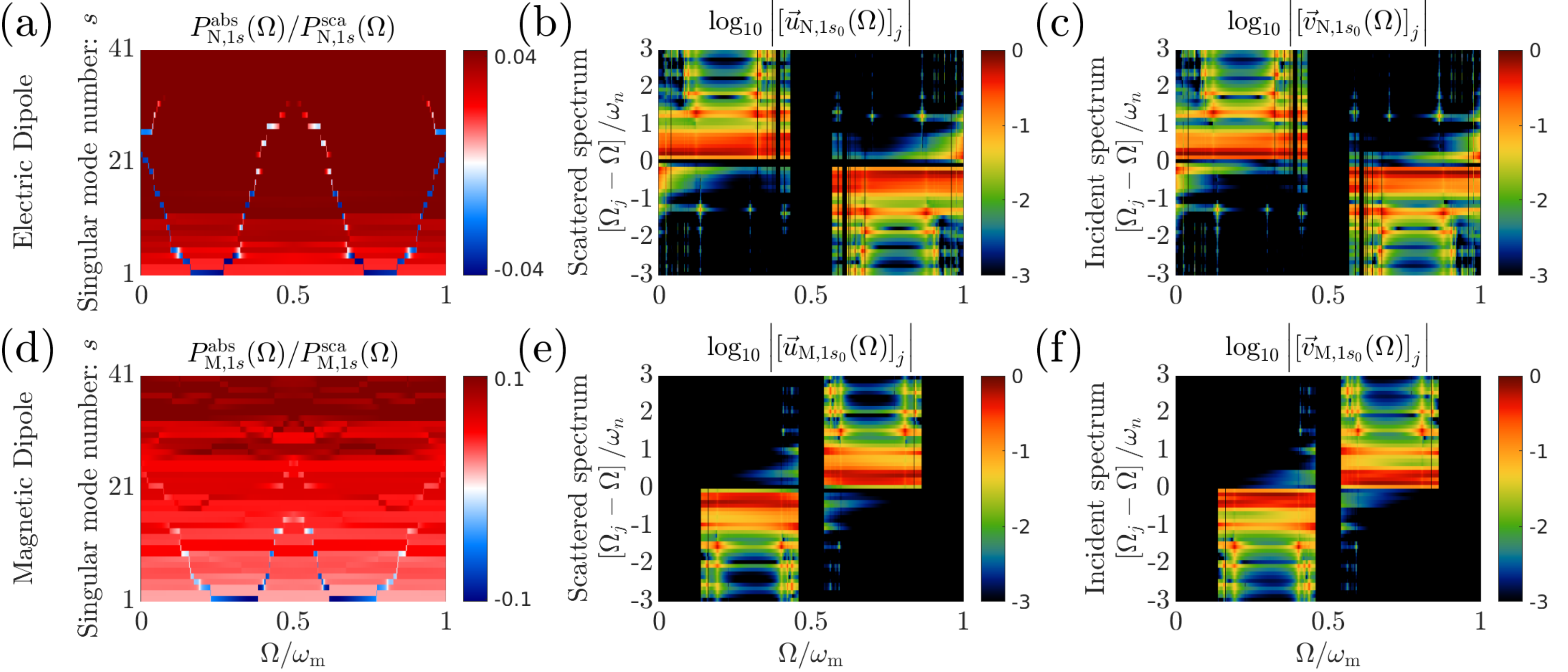}
	\caption{Singular value decomposition of the matrices of the scattering system revealing the presence of singular dipolar modes that demonstrate negative absorption, i.e., a transfer of energy from the time-varying matter to the photons in an inelastic scattering process. An optimized excitation of the system can give birth to such modes. (a,b) Plots of the ratio of the absorbed and scattered powers under the system's excitation with its singular modes. The presence of singular modes with negative absorption can be observed. Plots of the output (b,e) and input (c,f) spectra of the left- and right-singular vectors of the singular modes that demonstrate maximally negative absorption for each Floquet frequency $\Omega$.} 
	\label{fig_4}
\end{figure*}
Next, we will study the properties of the scattering system of a homogeneous spherical scatterer made of the time-varying and dispersive medium that was studied in the previous subsection. We consider the same material dispersion as the one used in Fig.~\ref{fig_2}(a), and we consider that the bulk electron density of the material is again modulated harmonically with a modulation frequency of $\omega_{\mathrm{m}}=\omega_n/10$ and a modulation strength of $M_\mathrm{s}=0.9$, as it was the case in the example illustrated in  Figs.~\ref{fig_2}(d-f). We consider the radius of the sphere $R$ to be equal to the free-space wavelength that corresponds to the resonance frequency of the Lorentz oscillator, i.e., $R=2\pi c_0/\omega_n$.

In Fig.~\ref{fig_3}(a) we plot the Mie coefficients (up to the quadrupolar order) that correspond to such a sphere in the non-time-varying case of $M_\mathrm{s}=0$. At very low frequencies, we observe the Rayleigh scattering region where the Mie coefficients diminish in amplitude. At low frequencies, away from the resonance frequency of the Lorentz oscillator where the material losses are maximized, we note the appearance of sharp and densely packed multipolar resonances. On the other hand, at frequencies larger than the resonance frequency of the Lorentz oscillator, we have a material with negative dielectric permittivity [see Fig.~\ref{fig_2}(a)] that demonstrates modes of much lower quality factors that are spectrally well-separated. 

Next, we introduce time modulation to the material from which the sphere is made. We plot in Figs.~\ref{fig_3}(b-e) the absolute values of the elements of T-matrices of such a time-modulated sphere given by Eq.~(\ref{eq:Tmatsca}). The results are plotted in a logarithmic scale for multipoles up to the quadrupolar order. We combine the calculated results for the T-matrices of all the Floquet frequencies in a single plot. There, the $x$-axis corresponds to the frequency $\omega^{\mathrm{inc}}$ of the incident multipolar excitation. The $y$-axis corresponds to the scattered band order of   radiating multipoles. The output frequency $\omega^{\mathrm{sca}}$ of the radiated multipole at a scattering band order $b_\mathrm{o}$ is given by the formula $\omega^{\mathrm{sca}}=\omega^{\mathrm{inc}}+b_\mathrm{o}\omega_{\mathrm{m}}$. A zero scattering band order, i.e., $b_\mathrm{o}=0$, means that the frequencies of the incident and scattered multipoles are the same. Therefore, in the limit of low modulation strengths $M_\mathrm{s}\rightarrow0$, we shall have a predominant response solely at the zero scattering band order, $b_\mathrm{o}=0$. The color of the plots encodes the amplitude of a radiated multipole at frequency $\omega^{\mathrm{sca}}$ once the sphere is excited by a single multipole of unit amplitude at frequency $\omega^{\mathrm{inc}}$. The white lines in Figs.~\ref{fig_3}(b-e) denote an  output frequency being zero, i.e., $\omega^{\mathrm{sca}}=0$. 

There are several interesting features to be observed in Figs.~\ref{fig_3}(b-e). Most importantly, we see that, due to the time modulation of the sphere's material, a monochromatic excitation gives rise to a polychromatic response. This implies an inelastic scattering process. The general condition for a resonant response is that the sphere is at resonance simultaneously both at the input and output frequencies $\omega^{\mathrm{inc}}$ and $\omega^{\mathrm{sca}}$. This, of course, happens predominantly when the input and output frequencies coincide. However, there are several other cases where such a resonant inelastic scattering process takes place. For example, we can see that there is a strong response along lines parallel to the white ones, where we have a constant output frequency $\omega^{\mathrm{sca}}$ that shall be associated with a sharp multipolar resonance supported by the sphere there. Such sharp resonances have a significant spectral echo predominantly in negative scattering band orders, with the response, though, weakening as the difference between the input and output frequencies increases. Moreover, we also observe the appearance of sharper features with an even stronger response along those spectral lines. We can associate these features with the simultaneous presence of sharp multipolar resonances at the respective input frequencies, leading to enhanced double-resonant effects. Furthermore, there is a beating pattern along those spectral lines. The periodicity thereof is related to the modulation frequency $\omega_{\mathrm{m}}$ and it indicates a spectral echo of a multipolar resonance at the input frequency. Another interesting feature is that even some coupling between input and output frequencies of opposite sign can be observed. This may lead to interesting phenomena such as parametric amplification~\cite{shcherbakov2019time,li2019beyond} and non-reciprocity~\cite{Our2}. Finally, we observe that for low input frequencies, the response of the sphere is weak, especially for larger multipolar orders, since in this case the optical size of the sphere is small.

As we highlighted already, there is an inelastic scattering process when we introduce a time modulation of the scatterer. It implies that the photons interact with the time-varying matter and exchange energy. This makes us wonder whether it is possible to create an active element out of such a time-varying scatterer that extracts energy from the time-varying matter and provides it to the photons. Therefore, we search for the possibility of using our scattering system to realize negative total absorption, even though the dispersive model of the Lorentz oscillator that we employ is rather lossy around the resonant frequency of the oscillator. Such an observation has already been reported in Ref.~\onlinecite{Stefanou:21} for a lossless system without material dispersion.

To this end, we perform a singular value decomposition of the T-matrices that correspond to each spectral comb with a specific Floquet frequency. The decomposition is given by Eq.~(\ref{eq:SVD}). The total scattered and absorbed energies by our scattering system are then given by Eqs.~(\ref{eq:Wsca2},\ref{eq:Wabs2}). Then, we excite our scattering system with each right-singular vector of the decomposed matrices, i.e., we consider excitations with $\mathcal{S}^{\mathrm{inc}}_{\alpha,\mu\nu s}(\Omega)=\delta_{\alpha\alpha'}\delta_{\mu\mu'}\delta_{\nu\nu'}\delta_{ss'}\delta(\Omega-\Omega')$ sweeping the values of $\alpha',s',\Omega'$, with $\mu'$ being arbitrary and $\nu'$ fixed to 1 as we focus on the dipolar response of the system. For all such excitations of our system, we observe the sign of the absorbed power. Exciting the system with a single right-singular vector means that we excite only a single spectral comb of some Floquet frequency, with a particular spectral distribution of the power over the frequencies of the comb. Simultaneously, our excitation consists of a single incoming multipole (dipole). Therefore, it is a quite special excitation, not only spectrally but also spatially. It corresponds to an excitation with a particular angular spectrum of plane waves that comprise such an incoming multipole. An arbitrary excitation of the system can be decomposed into this basis of right-singular vectors. Exciting our system, though, with a single singular mode enables us to ignore the inter-modal couplings due to the terms in the third row of Eq.~(\ref{eq:Wabs2}).

We consider as a scattering system the same sphere that we studied before in this subsection, and we plot our results in Fig.~\ref{fig_4}. The singular modes are ordered in a descending order of their respective singular values, i.e., in a descending order of total scattered power,  as it is implied by Eq.~(\ref{eq:Wsca2}). In Figs.~\ref{fig_4}(a,d), we observe that for many of the spectral combs with varying Floquet frequency $\Omega$, we can have singular modes that demonstrate significantly negative values of absorbed power, i.e., a significant transfer of energy from the time-varying matter to the photons of the electromagnetic field during the inelastic scattering process. Such modes can be excited only with the particular illumination of the corresponding right-singular vectors. The spectral content of the right-singular vectors, $\vec{{v}}_{\alpha,1s_0}(\Omega)$, which correspond to singular modes that demonstrate a maximally negative absorption (indicated with the index $s_0$), is plotted in logarithmic scale and for each Floquet frequency in Figs.~\ref{fig_4}(c,f). The black-colored columns of the figure indicate an absence of a singular mode with negative absorption for that particular Floquet frequency. In Figs.~\ref{fig_4}(b,e) we plot the norm of the elements of the corresponding left-singular vectors $\vec{{u}}_{\alpha,1s_0}(\Omega)$, i.e., the spectral content of the scattered fields once the system gets excited by the corresponding right-singular vectors. We observe that the input and output spectra of the singular modes that demonstrate negative absorption are characterized by a spectral distribution of power predominantly over the low frequencies where the material losses due to dispersion are low. Due to the presence of a lossy spectral region and the size of the considered sphere, we do not find any singular mode with negative absorption for the quadrupolar modes. It would only become possible for larger sizes of the sphere. It is rather remarkable that negative absorption can be achieved even in the presence of strong material losses once we optimize the system's excitation. Finally, let us note that the presence of the third row of Eq.~(\ref{eq:Wabs2}), corresponding to inter-modal couplings, allows for the possibility of attaining negative absorption under other excitation schemes as well, that involve, in general, a superposition of such singular modes.
\subsection{Numerical performance of the developed algorithm in comparison to a full-wave solver}
\begin{figure*}
	\includegraphics[width=17cm]{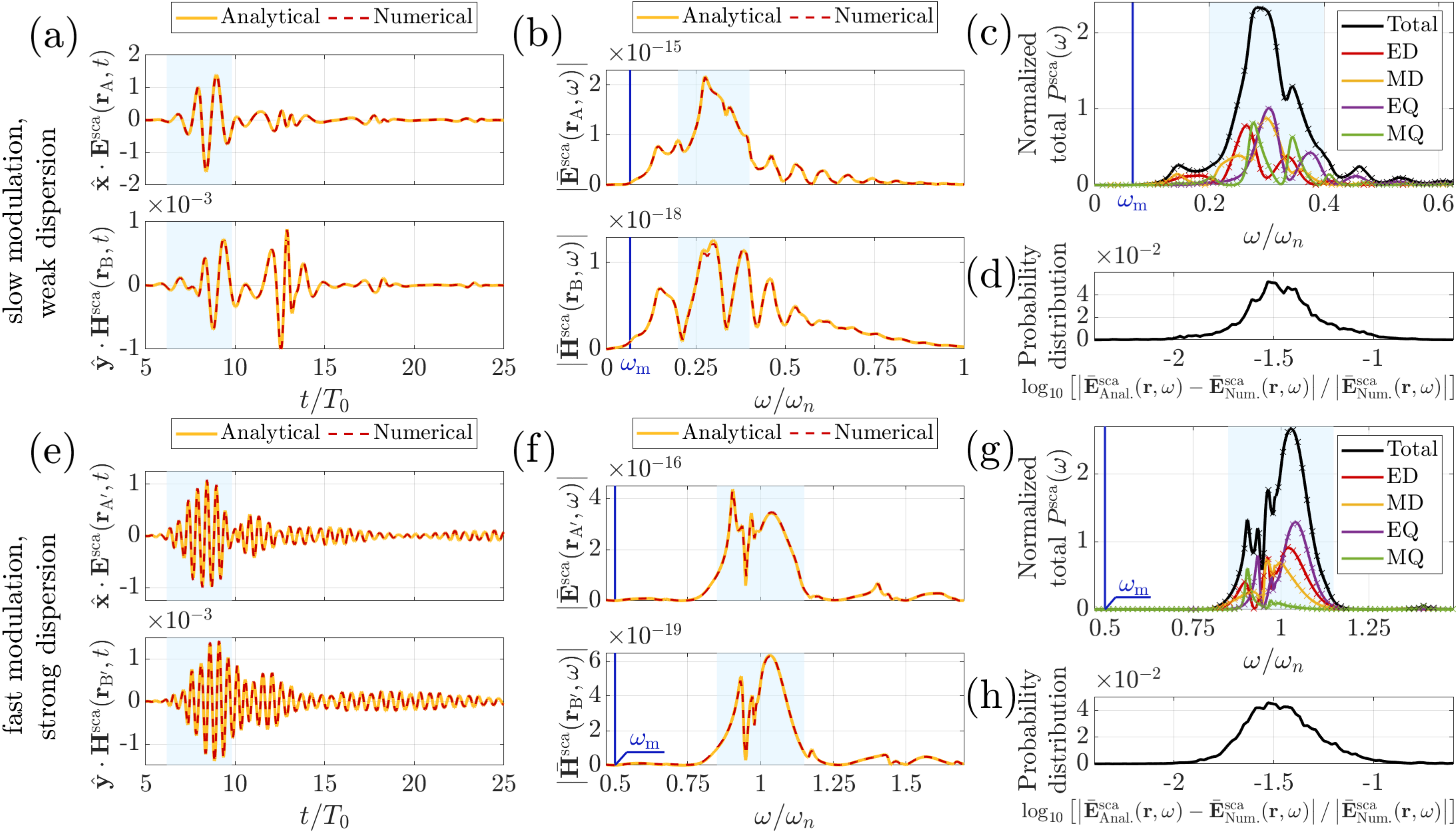}
	\caption{Comparison of analytical and numerical results from a full-wave time-domain solver, for two sets of simulations: (a-d) ``slow modulation, weak dispersion'' and (e-h) ``fast modulation, strong dispersion''. (a,e) Plots of the electric (at points A and A', respectively) and magnetic (at points B and B', respectively) fields in the time domain. (b,f) Plots of the Fourier transform of the fields in (a,e). The highlighted light-blue regions indicate the temporal/spectral windows where 99$\%$ of the energy of the incident Gaussian pulse resides. (c,g) Plots of the normalized total scattered power together with the individual multipolar contributions. Solid lines show the analytical results, whereas cross-markers indicate the numerical results of the full-wave simulation. (d,h) Plots of the probability distribution of the logarithmic relative error between the fields calculated analytically and numerically. These plots involve the error statistics among a considered distribution of points over a spherical shell surrounding the scatterer, and, also, over a broadband spectral window.} 
	\label{fig_5}
\end{figure*}
Our numerical analysis of the problem of scattering by a time-varying and dispersive sphere was verified by full-wave simulations performed in Comsol Multiphysics with the finite element time-domain method. To solidify the comparison and study the efficiency of the developed semi-analytical approach, we have considered two different sets of simulations. We again adopt the Lorentz oscillator model with time-modulated bulk electron density to account for material dispersion and time modulation. Equation~(\ref{eq:difEqPolDens}) is embedded inside the full-wave numerical solver. We name the first set as ``slow modulation, weak dispersion'' since we consider a relatively slow but still strong modulation of the medium of the scatterer. The scatterer is excited in a spectral window characterized by  weak material dispersion, i.e., far away from the resonance of the Lorentz oscillator. On the other hand, the second set of simulations also considers a strong modulation amplitude, but now with a fast modulation frequency. Moreover, the sphere is excited in a spectral window centered around the resonance frequency of the Lorentz oscillator, where we encounter maximal dispersion. Hence, we name this second set of simulations as ``fast modulation, strong dispersion''.

For the first set of simulations, we consider a Lorentz model with damping factor $\gamma_{n}=\omega_{n}/8$ and bulk electron density $N_0=11\omega_n^2m_e\varepsilon_0/e^2$. This material can be considered dispersionless far away from its natural frequency and, therefore, we simulate the excitation [see Eq.~(\ref{eq:Einctime})] of the sphere with a Gaussian pulse of unit amplitude $E_0=1$ V/m and temporal width $T_0=2.9\times2\pi/\omega_{n}$ that is centered at the frequency $\omega_0=0.3\omega_{n}$. The pulse is polarized along the $x$-axis and propagates along the $+z$-direction
~\footnote{Let us note that while for our analytical calculations we use an infinitely extended plane wave, for the Comsol simulations we use a Gaussian beam with an optically large waist to approximate the plane-wave excitation in our numerical setup.}. The pulse is temporally centered at $t_0=8\,T_0$. The material is modulated with frequency $\omega_{\rm m}=\omega_{n}/15$ and modulation strength $M_{\rm s}=0.9$. We choose a sphere radius of $R=7.095\,c_0/\omega_n$ to ensure a significant scattering response. This combination of excitation, modulation parameters, and sphere dimensions provides a rich scattering spectrum. Locating the sphere at the origin of the coordinate system, we compare the fields at two arbitrarily chosen spatial points: behind [point A(0,0,$1.43R$)] and above [point B($1.43R$,0,0)] the sphere. Since the electric fields' $x$-component is the strongest at point A and the magnetic fields' $y$-component is strong at point B, we present them in time domain in Fig.~\ref{fig_5}(a). The respective norms of the fields are presented in frequency domain in Fig.~\ref{fig_5}(b). The semi-analytical results obtained from the developed theory are plotted with solid yellow lines. The numerical results from Comsol are plotted with dashed red lines. The highlighted light-blue area denotes the temporal/spectral region where 99$\%$ of the energy of the incident pulse resides. We observe that the results obtained by the numerical simulations almost ideally match the semi-analytical results. 

In Fig.~\ref{fig_5}(c), we plot the total power scattered by the sphere [see Eq.~(\ref{eq:Wsca1})], i.e., the power that we calculate by integrating the power flux of the Poynting vector of the scattered field over a spherical shell that surrounds the scatterer. To perform a multipolar decomposition of the scattered fields numerically, we measure the fields over a distribution of points located over a spherical shell of radius $1.43R$. We use a surface integral (see Eq.~(5.175) from Ref.~\onlinecite{mishchenko2002scattering}) to extract from the Comsol simulations the multipolar amplitudes of the scattered field $\mathcal{A}_{\alpha,\mu\nu}^\mathrm{sca}(\omega)$ and compare them with the ones that we obtain analytically. The scattered power spectra are normalized to the spectral peak of the total power flux of the incident field passing through the geometrical cross-section of the scatterer. The values of the total scattered power are plotted with a black solid line. The individual multipolar contributions (up to the quadrupolar order) are plotted with colored solid lines. We use cross-markers to plot the numerical results obtained from Comsol. Again, we see a perfect agreement between the semi-analytical and the numerical results. Finally, in Fig.~\ref{fig_5}(d), we plot the probability distribution of the logarithmic relative error between the analytically and numerically calculated fields over the points of the previously considered spherical shell surrounding the scatterer and over a spectral window between the frequencies $[0.1\omega_n,0.93\omega_n]$, where the signals are strong. The graph indicates a relative error distribution predominantly within the range of 1$\%$ and 10$\%$.

The second set of simulations considers a less lossy material for which the Lorentz model parameters read $\gamma_{n}=\omega_n/120$ and $N_0=1.12\omega_n^2m_e\varepsilon_0/e^2$. To study the effects of strong dispersion, we excite the scatterer at the resonance frequency (i.e., $\omega_0=\omega_{n}$). To capture a rich frequency spectrum, we choose the pulse width $T_0=1.934\times2\pi/\omega_{n}$ and the sphere radius $R=1.824\, c_0/\omega_n$. The pulse is again temporally centered at $t_0=8\,T_0$. The modulation strength of the material is considered again to be $M_{\rm s}=0.9$. In contrast, the modulation frequency is now $\omega_{\rm m}=\omega_{n}/2$, which corresponds to a relatively high modulation speed. Figures~\ref{fig_5}(e-h) are the counterparts to Figs.~\ref{fig_5}(a-d), but now for the case of the second set of simulations. The only difference is that in Figs.~\ref{fig_5}(e,f) the observation points are located at A$'$(0,0,$2.432R$) and B$'$($2.432R$,0,0). As for the Figs.~\ref{fig_5}(g,h), we have the observation points located over a spherical shell of radius $2.432R$, and the spectral window considered for the statistics of Fig.~\ref{fig_5}(h) being between the frequencies $[0.827\omega_n,1.172\omega_n]$. Again, we can observe that for the second set of simulations characterized by fast modulation and strong dispersion, the simulation and analytical results are in almost perfect agreement. 

In Appendix~\ref{app:app3} we present results for a comparative study between two different material models: one accounting for temporal dispersion, as it was the case with the results presented in this subsection, and the other ignoring it. Our results there propound the appreciation of the importance of taking into account the temporal dispersion.

Finally, let us highlight that the numerical simulations are computationally considerably more demanding than the presented semi-analytical approach. While the Comsol simulations for the first setup lasted for 12 days requiring 110 gigabytes of RAM, and for the second setup they lasted for 5 days requiring 43 gigabytes of RAM, the semi-analytical algorithm uses 2 gigabytes of RAM to calculate T-matrices and only needs approximately 15 seconds for both setups.
\section{Conclusion}
To summarize, we have presented an analytical model that describes light scattering on spheres made of dispersive and time-varying media. First, we comprehensively studied the propagation of electromagnetic waves in unbounded time-varying media with frequency dispersion. We then applied this theory to treat the problem of light scattering by spheres composed of such media. In contrast to other approaches for theoretical investigations of  time-modulated structures, the developed route considers spatially confined scatterers, incorporates frequency dispersion, and allows an arbitrary modulation speed and amplitude. In addition to that, we verify our findings using full-wave simulations. 

This study can be considered referential since it treats such a canonical object as a sphere. It makes an essential initial step towards a general understanding of all kinds of scattering effects in time-varying structures. It has the crucial advantage that it can be used to study all kinds of effects considering a simple shaped object such as a sphere in a short amount of time and with minimal computational resources. The understanding borne from these  investigations provides the language to discuss more elaborate systems that are no longer feasible for an analytical treatment but require a numerical full-wave simulation to capture all the details. In the past, the analytical solution of the canonical problem of scattering by a stationary sphere was one of the key cornerstones in the development of the theory of light scattering, and we hope that this extension of the theory to time-varying canonical scatterers will serve the same important purpose.

A further extension of this study can include a proper description of absorption with the associated dispersion. Such a study would be crucial for providing insights into parametric amplification for realistic systems. Alternatively, this study can be further extended towards non-spherical geometries using, for instance, an analytical solution for other simple structures, such as slabs or cylinders, or employing simulations for more complicated three-dimensional structures. In addition to that, one can consider arrays of time-varying particles.  
\begin{acknowledgments}
 We acknowledge support by the German Research Foundation through Germany’s Excellence Strategy via the Excellence Cluster 3D Matter Made to Order (EXC-2082/1 - 390761711). A. G. L. acknowledges support from the Max Planck School of Photonics, which is supported by BMBF, Max Planck Society, and Fraunhofer Society and from the Karlsruhe School of Optics and Photonics (KSOP). T. K. acknowledges support from the Alexander von Humboldt Foundation through the Humboldt Research Fellowship for postdoctoral researchers. R. A. acknowledges support from the Alexander von Humboldt Foundation through the Feodor Lynen (Return) Research Fellowship. G. P. acknowledges partial funding from the Academy of Finland (project 330260). V. A. and S. F. acknowledge the support of a MURI project from the U.S. Air Force of Office of Scientific Research (Grant No FA9550-21-1-0244).
\end{acknowledgments}

\bibliography{references}

\appendix

\section{Vector Spherical Harmonics (VSHs) }
\label{app:app1}
In this section, we will provide the basic definitions of the VSHs that play a central role in our theoretical analysis in the main text. We begin with defining the scalar spherical harmonics, $\psi^{(\iota)}_{\mu\nu\kappa}(\mathbf{r})$, i.e., the spherical solutions of the scalar monochromatic Helmholtz equation. Their spatial representation in the spherical coordinate system is given by the formula~\cite{morse1953methods}:
\begin{eqnarray}
\psi^{(\iota)}_{\mu\nu\kappa}(\mathbf{r})&=&\gamma_{\mu\nu}\hspace{2pt} z^{(\iota)}_{\mathrm{M},\nu}\left(\kappa \mathrm{r} \right)\mathrm{P}_\nu^\mu(\mathrm{cos}\theta) e^{\mathrm{i}\mu\phi},\label{eq:sscalarwave}
\end{eqnarray}
where $\kappa$ is the wavenumber of the monochromatic wave equation, $\nu(\nu+1)$ is the eigenvalue of the squared total angular momentum operator $\mathbf{J}^2$ with $\nu$ taking integer values $1,2,\dots$ and  $\mu$ is the eigenvalue of the projection of the total angular momentum operator along the $z$-axis $J_z$ taking integer values $-\nu,\dots,0,\dots,\nu$. $z^{(\iota)}_{\mathrm{M},\nu}\left(x\right)$ denotes the spherical Bessel ($\iota=1$) and Hankel ($\iota=3$) functions of the first kind of order $\nu$. $\mathrm{P}_\nu^\mu(x)$ are the associated Legendre functions of the first kind and $\gamma_{\mu\nu}=\sqrt{\frac{(2\nu+1)(\nu-\mu)!}{4\pi\nu(\nu+1)(\nu+\mu)!}}$ are normalization coefficients.

By following Ref.~\onlinecite{morse1953methods}, we can construct a full-set of divergence-free vector spherical harmonics based on the above scalar spherical harmonics. We will use the symbol $\alpha$ to denote the TE ($\alpha=\mathrm{M}$) and TM ($\alpha=\mathrm{N}$) eigenwaves. $\alpha$ can be associated with the eigenvalue of the parity operator, of which the VSHs are eigenstates. The spatial representation of such VSHs, $\mathbf{F}^{(\iota)}_{\alpha,\mu\nu\kappa}(\mathbf{r})$, is given by the formulas below ~\cite{morse1953methods,mishchenko2002scattering}:
\begin{subequations}
\begin{eqnarray}
\hspace{-8mm}\mathbf{F}^{(\iota)}_{\mathrm{M},\mu\nu\kappa}(\mathbf{r}) &\triangleq&\nabla \times\left[\mathbf{r}\psi^{(\iota)}_{\mu\nu\kappa}(\mathbf{r})\right]
= \mathrm{i}z_{\mathrm{M},\nu}^{(\iota)}(\kappa \mathrm{r} )\mathbf{f}_{\mathrm{M},\mu\nu}(\hat{ \mathbf{r} }),  \label{eq:sfairM}\\
\hspace{-8mm}\mathbf{F}^{(\iota)}_{\mathrm{N},\mu\nu\kappa}(\mathbf{r}) &\triangleq&\frac{1}{\kappa}\nabla \times\mathbf{F}^{(\iota)}_{\mathrm{M},\mu\nu\kappa}(\mathbf{r})\nonumber\\
&=& \hat{ \mathbf{r} }\frac{\nu(\nu+1)}{\kappa \mathrm{r} }\psi^{(\iota)}_{\mu\nu\kappa}(\mathbf{r}) + z_{\mathrm{N},\nu}^{(\iota)}(\kappa \mathrm{r} )\mathbf{f}_{\mathrm{N},\mu\nu}(\hat{ \mathbf{r} }),\hspace{6pt}\label{eq:sfairN}
\end{eqnarray}
\end{subequations}
where:
\begin{subequations}
\begin{eqnarray}
\mathbf{f}_{\mathrm{M},\mu\nu}(\hat{ \mathbf{r} }) &=& \gamma_{\mu\nu}\left[\hat{\theta}\tau_{\mu\nu}^{(1)}(\theta) + \mathrm{i}\hat{\phi}\tau_{\mu\nu}^{(2)}(\theta) \right]e^{\mathrm{i}\mu\phi},\label{eq:fsfairM}\\
\mathbf{f}_{\mathrm{N},\mu\nu}(\hat{ \mathbf{r} })&=&  \gamma_{\mu\nu}\left[\hat{\theta}\tau_{\mu\nu}^{(2)}(\theta) +  \mathrm{i}\hat{\phi}\tau_{\mu\nu}^{(1)}(\theta) \right]e^{\mathrm{i}\mu\phi}, \label{eq:fsfairN}\\
\tau_{\mu\nu}^{(1)}(\theta) &=& \mu\frac{\mathrm{P}_\nu^\mu(\mathrm{cos}\theta)}{\mathrm{sin}\theta},\label{eq:leg1}\\
\tau_{\mu\nu}^{(2)}(\theta) &=&\frac{\partial \mathrm{P}_\nu^\mu(\mathrm{cos}\theta)}{\partial\theta},\label{eq:leg2}
\end{eqnarray}
\text{and $z_{\mathrm{N},\nu}^{(\iota)}(x)$ is defined as:}
\begin{eqnarray}
z_{\mathrm{N},\nu}^{(\iota)}(x)&=&\frac{1}{x}\frac{\partial}{\partial x}[x \hspace{2pt}z_{\mathrm{M},\nu}^{(\iota)}(x)].\label{eq:riccati}
\end{eqnarray}
\end{subequations}
\section{Expansion of the incident field in series of VSHs}
\label{app:app2}
In this section, we are discussing the coefficients $\mathcal{A}^{\mathrm{inc}}_{\alpha,\mu\nu}(\omega)$ that expand the incident field in series of VSHs. Explicit expressions for these coefficients, for the cases of incoming focused beams or even single propagating plane waves can be found in Ref.~\onlinecite{Lamprianidis2018}. Moreover, the translation addition theorem of the radiating VSHs [see Eq.~(C.68) in Ref.~\onlinecite{mishchenko2002scattering}] can provide expressions for the coefficients for the case of multipolar emitters placed in the vicinity of the scatterer. 

In this work, for the numerical demonstration of our algorithm, we have been considering an excitation by an $x$-polarized plane wave propagating along the $z$-axis and having a Gaussian pulse envelope with width $T_0$ and carrier frequency $\omega_0$. Its representation in time domain is given by the formula
\begin{eqnarray}
\hspace{-7mm}\mathbf{E}^{\mathrm{inc}}(\mathbf{r},t)\hspace{-0.7mm}&=&\hspace{-0.7mm}E_0\hat{\mathbf{x}}\hspace{1pt}e^{-\frac{(t-t_0-z/c_0)^2}{2T_0^2}}\mathrm{cos}[\omega_0(t-t_0-z/c_0)],\hspace{0pt}\label{eq:Einctime}
\end{eqnarray}
whereas its representation in the frequency domain is given by
\begin{eqnarray}
\overline{\mathbf{E}}^{\mathrm{inc}}(\mathbf{r},\omega)&=&\hat{\mathbf{x}}\,e^{\mathrm{i}\omega z/c_0}\nonumber\\
\times&& \hspace{-12pt}\frac{E_0T_0}{2}\left[e^{-\frac{T_0^2(\omega_0-\omega)^2}{2}}+e^{-\frac{T_0^2(\omega_0+\omega)^2}{2}}\right]e^{\mathrm{i}\omega t_0},\hspace{15pt}
\label{eq:Eincfrec}
\end{eqnarray}
where $t_0$ is a time delay and $c_0$ is the speed of light in free space. Expanding the plane wave $\hat{\mathbf{x}}e^{\mathrm{i}\omega z/c_0}$ in a series of VSHs around the origin of the coordinate system [see Eq.~(\ref{eq:EincExpans}) here and Eq.~(8) in Ref.~\onlinecite{Lamprianidis2018}], we finally get the following expression for the incident spherical amplitudes:
\begin{eqnarray}
\mathcal{A}^{\mathrm{inc}}_{\alpha,\mu\nu}(\omega)&=&E_0T_0\pi\hspace{1pt} \mathrm{i}^{\nu+1}\hspace{1pt} \gamma_{-1 \nu}\left[\delta_{\mu,1}+(-1)^{\delta_{\alpha N}}\delta_{\mu,-1}\right]\nonumber\\
&&\left[e^{-\frac{T_0^2(\omega_0-\omega)^2}{2}}+e^{-\frac{T_0^2(\omega_0+\omega)^2}{2}}\right]e^{\mathrm{i}\omega t_0},\label{eq:Eincampl}
\end{eqnarray}
where $\delta_{ij}$ is the Kronecker delta. Let us note that the above equation holds true only for positive frequencies $\omega$. For the negative frequencies we can make use of the following symmetry property that satisfies the condition for real fields in time domain: {$\mathcal{A}^{\mathrm{inc}}_{\alpha,\mu\nu}(-\omega)=(-1)^{\nu+\mu+\delta_{\alpha N}}\left[\mathcal{A}^{\mathrm{inc}}_{\alpha,-\mu\nu}(\omega)\right]^*.$} 
\section{The compromise effects of ignoring temporal dispersion}
\label{app:app3}
\begin{figure*}
	\includegraphics[width=17cm]{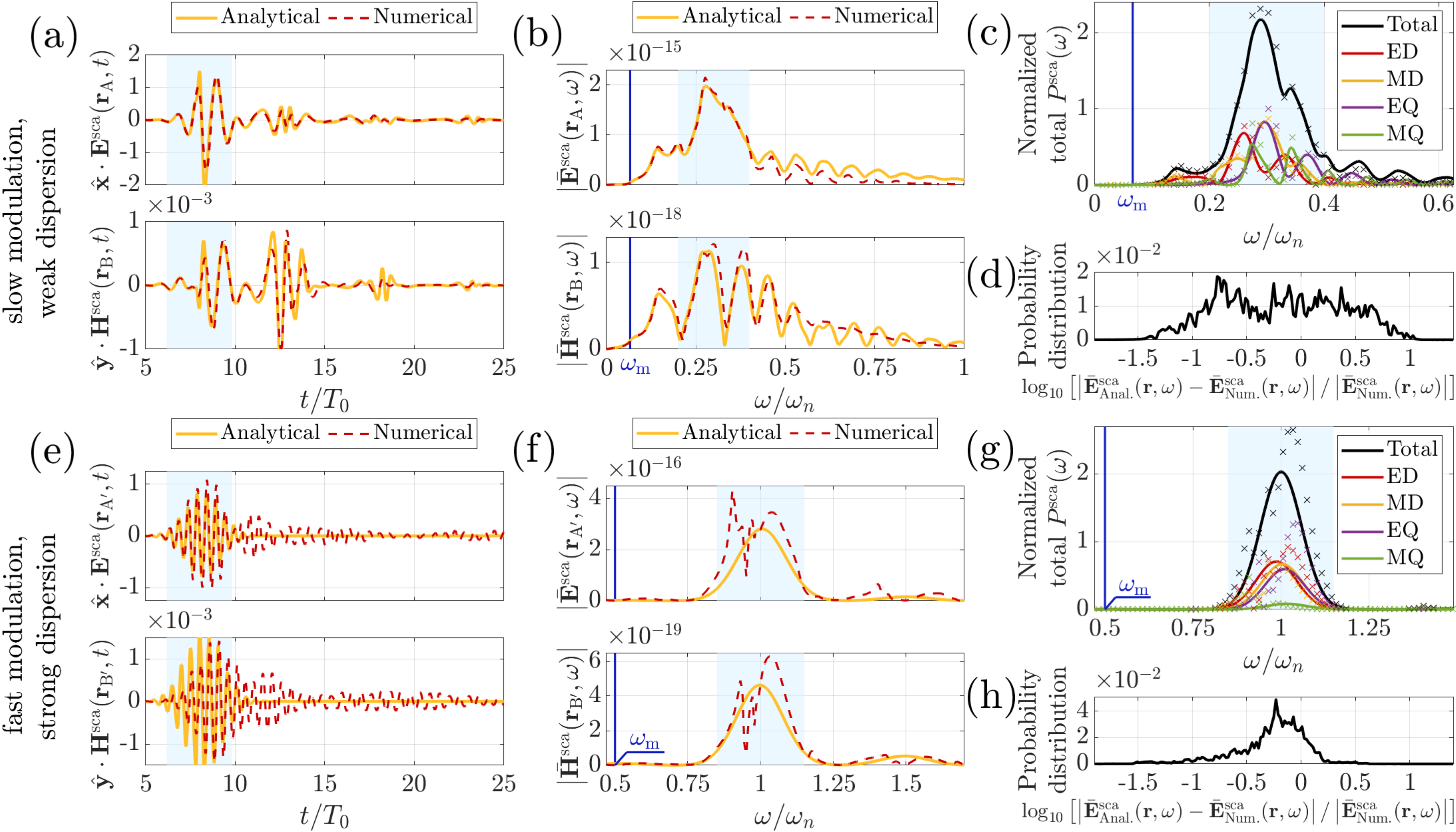}
	\caption{Same as Fig.~\ref{fig_5} but with the ``Analytical" results being based on a material model that ignores temporal dispersion. The effects of the compromise of ignoring temporal dispersion are clearly visible, especially in the "strong dispersion" case.} 
	\label{fig_5App}
\end{figure*}
In this section, we study the impact of ignoring temporal dispersion in our material model on the scattering response from the spheres. The results are shown in Fig.~\ref{fig_5App}, where the same cases as previously studied are considered with the notable difference that the ``Analytical" results correspond here to a material model with no temporal dispersion. Specifically, the electric response function of the time-varying medium, instead of being given by Eq.~(\ref{eq:respfunc}), as it was the case in Fig.~\ref{fig_5}, here it is given by:
\begin{eqnarray}
\overline{R}_\mathrm{e}(\omega-\omega',\omega')&=&\frac{1}{\sqrt{2\pi}}\frac{e^2}{m_e\varepsilon_0}\hspace{2pt}\frac{\overline{N}(\omega-\omega')}{\omega_n^2-\omega_0^2-\mathrm{i}\gamma_n\omega_0}.\label{eq:respfuncnodisp}
\end{eqnarray}
This model considers a vanishing dispersion across all frequencies, i.e. a material constant that takes the value equal to what we had previously at the frequency $\omega_0$, the central frequency of the incident pulse. Also, let us clarify that the ``Numerical" results here are the same as those of Fig.~\ref{fig_5}, i.e., they correspond to full-wave simulations that account for temporal dispersion.

We can observe that disregarding temporal dispersion in our material model can lead to significant deviations. The deviations are more pronounced in the ``strong dispersion" case, where we excite our sphere within a spectral region where the time-varying medium is characterized by strong temporal dispersion. The importance of accounting for material temporal dispersion is highlighted by our results here.

\end{document}